\documentclass{CSML}

\usepackage{lastpage}
\def\dOi{13(3:2)2017}
\lmcsheading%
{\dOi}
{1--\pageref{LastPage}}
{}
{}
{Dec.~24, 2014}
{Jul.~\phantom06, 2017}
{}

\usepackage{hyperref}\hypersetup{hidelinks}

\usepackage{multicol}
\usepackage{amsmath}
\usepackage{proof}
\usepackage{stmaryrd}
\usepackage{latexsym}
\usepackage{amssymb}
\usepackage{amsfonts}
\usepackage{esvect}
\usepackage{alltt}
\usepackage{bussproofs}
\usepackage[all]{xy}
\usepackage{prettyref}
\usepackage{tabularx}
\usepackage{adjustbox}

\newcommand{\hideshow}[1]{{\mbox{}}}

\input{macros.sty}

\begin{document}

\title[\LLFP : A Logical Framework using Monads]{\LLFP : A Logical Framework for modeling External Evidence,
  Side Conditions, and Proof Irrelevance using Monads}
\author[F.~Honsell]{Furio Honsell\rsuper a} 
\address{{\lsuper{a,d}}Universit\`{a} di Udine, Italy} 
\email{\{furio.honsell, ivan.scagnetto\}@uniud.it} 

\author[L.~Liquori]{Luigi Liquori\rsuper a} 
\address{{\lsuper a}Inria, France} 
\email{luigi.liquori@inria.fr} 

\author[P.~Maksimovi\'{c}]{Petar Maksimovi\'{c}\rsuper c} 
\address{{\lsuper c}Inria, France and \newline Mathematical Institute of the
  Serbian Academy of Sciences and Arts, Serbia}
\email{petarmax@mi.sanu.ac.rs} 

\author[I.~Scagnetto]{Ivan Scagnetto\rsuper d} 
\address{\vspace{-18 pt}} 


\begin{abstract}
  We extend the constructive dependent type theory of the Logical
  Framework \LF\ with \emph{monadic}, \emph{dependent} type
  constructors indexed with predicates over judgements, called
  \emph{Locks}. These monads capture various possible \emph{proof
    attitudes} in establishing the judgment of the object logic
  encoded by an \LF\ type. Standard examples are \emph{factoring-out}
  the verification of a \emph{constraint} or \emph{delegating} it to
  an \emph{external oracle}, or supplying some \emph{non-apodictic}
  epistemic evidence, or simply \emph{discarding} the proof witness of
  a precondition deeming it irrelevant.  This new framework, called
  \emph{Lax Logical Framework}, \LLFP, is a conservative extension of
  \LF, and hence it is the appropriate metalanguage for dealing
  formally with \emph{side-conditions} in rules or \emph{external
    evidence} in logical systems.  \LLFP\ arises once the
  \emph{monadic} nature of the \emph{lock} type-constructor,
  $\Lock{\P}{M}{\sigma}{\cdot}$, introduced by the authors in a series
  of papers, together with Marina Lenisa, is fully exploited. The
  nature of the lock monads permits to utilize the very \emph{Lock}
  destructor, $\Unlock{\P}{M}{\sigma}{\cdot}$, in place of Moggi's
  monadic $let_T$, thus simplifying the equational theory. The rules
  for $\Unlock{\P}{M}{\sigma}{\cdot}$ permit also the removal of the monad
  once the constraint is satisfied. We derive the meta-theory of
  \LLFP\ by a novel indirect method based on the encoding of \LLFP\ in
  \LF. We discuss encodings in \LLFP\ of call-by-value
  $\lambda$-calculi, Hoare's Logic, and Fitch-Prawitz Naive Set
  Theory. 
\end{abstract}

\maketitle

\begin{flushright} \emph{dedicated to Pierre-Louis
    Curien}
\end{flushright}
\vfill\newpage

\tableofcontents

\section{Introduction}
\begin{flushright}
  \emph{A mathematician, half way through a proof, during a seminar,
    said ``...and this trivially holds''.  But after a few seconds of
    silence, somewhat to himself, he mumbled: ``\ldots but is this
    really trivial, here? \ldots Hmm \ldots''.  He kept silent for 5
    minutes.  And finally triumphantly exclaimed ``Yes, it is indeed
    trivial!''}
\end{flushright}

\bigskip

In this paper we introduce, develop the metatheory, and give
applications of the \emph{Lax Logical Framework}, \LLFP. \LLFP\ is a
conservative extension of \LF\ which was first outlined in the
Symposium in honour of Pierre Louis Curien, held in Venice in September
2013. A preliminary version of \LLFP\ was presented in
\cite{llfp-mfcs2014}.  This system has grown out of a series of papers on
extensions of \LF\ published by the authors, together with Marina
Lenisa, in recent years
\cite{BCKL-03,HLL06,HLLS08,honsell:hal-00906391,Honsell:2013:YFP:2503887.2503896}. The
idea underpinning these systems is to be able to express explicitly,
by means of a new type-constructor $\Lock{\P}{M}{\sigma}{\cdot}$,
called a \emph{lock}, the fact that in order to obtain a term of a given
type it is necessary to verify the constraint
${\P}(\Gamma \VDASHS M : \sigma)$. By using this type constructor, one can
capture various \emph{proof attitudes} which arise in practice, such
as \emph{factoring-out} or \emph{postponing} the verification of
certain judgements whose evidence we do not want to derive in the
standard way. This occurs when the evidence for the justification of
that judgement is supplied by an \emph{external proof search tool} or
an \emph{external oracle}, or some other \emph{non-apodictic}
epistemic sources of evidence such as diagrams, physical analogies, or
explicit computations according to the \emph{Poincar\'e Principle}
\cite{bar02}. These proof attitudes are ultimately similar to that
which occurs in proof irrelevant approaches when one is only
interested to know that some evidence is there, but the precise nature
of the proof witness is declared immaterial. Therefore, locked types
allow for a straightforward accommodation within the Logical Framework
of many different \emph{proof cultures} that otherwise can be embedded
only very deeply~\cite{deep,hirschkoff:bisimproofs} or axiomatically
\cite{HMS-01}. Locked types support the main motivation of \LLFP,
namely that external tools may be \emph{invoked} and \emph{recorded}
uniformly in an \LF\ type-theoretic framework.

The main novelty of \LLFP\ \wrt\ previous systems using locked types
introduced by the authors, is that \LLFP\ capitalizes on a monadic understanding of
$\Lock{\P}{M}{\sigma}{\cdot}$ constructors. An extended abstract of the present paper appears in~\cite{llfp-mfcs2014}\footnote{The version of \LLFP\ introduced here is both a
restriction and an \emph{errata corrigenda} of the system in~\cite{llfp-mfcs2014}.
 The present system is a restriction w.r.t.~\cite{llfp-mfcs2014}, in that the assumptions of the $(O{\cdot}Guarded{\cdot}Unlock)$ rule are less general than the one in~\cite{llfp-mfcs2014}, but it is an \emph{errata corrigenda} in that the present rule is slightly rephrased and the new rule $(F{\cdot}Guarded{\cdot}Unlock)$ is introduced, so as to allow one to prove the subject-reduction without any assumptions. Hence we discard the system
in~\cite{llfp-mfcs2014}, and replace it by the present one even in~\cite{llfp-mfcs2014}. We call, therefore, the system in the present paper
 \emph{the} Lax Logical Framework, even if this name was already
used for the one in~\cite{llfp-mfcs2014}. Signatures and derivations discussed in~\cite{llfp-mfcs2014} carry through in the present version ``as is''.}. Hence, \LLFP\ can be viewed
as the extension of \LF\ with a family of \emph{monads} indexed with
predicates over typed terms, which capture the \emph{effect} of
factoring out, or postponing, or delegating to an external oracle the
task of providing a proof witness of the verification of the
\emph{side-condition} ${\P}(\Gamma \VDASHS M : \sigma )$.  The basic
idea is that any constraint ${\P}$ can be viewed as a monad
$T_\P$. Its natural transformation
$\eta_{T_\P}: A \rightarrow T_\P(A) $ amounts to a sort of
\emph{weakening}, namely any judgement can always be asserted subject
to the satisfaction of a given constraint. Correspondingly, the other
canonical natural transformation
$\mu_{T_\P}: T^2_\P(A) \rightarrow T_\P(A)$, amounts to a sort of
\emph{contraction}, corresponding to the fact that we trust the
verifier, and hence verifying a given constraint twice is redundant.

Being a conservative extension of \LF, \LLFP\ can be used as a
metalanguage for defining logics and proofs. Furthermore, \LLFP\
can be used as a \emph{platform} for checking proof arguments that
combine different systems or invoke external oracles. Correctness of
proofs in \LLFP\ is, therefore, \emph{conditionally decidable}, \ie\ it is
decidable provided the external predicate is decidable.

Following the paradigm of Constructive Type Theory, once the new
locked type constructor is introduced, we introduce also the
corresponding lock constructor for \emph{terms}, which we continue to
denote as $\Lock{\P}{M}{\sigma}{\cdot}$, together with the \emph{unlock}
destructor for terms $\Unlock{\P}{M}{\sigma}{\cdot}$. This latter
term constructor allows one to exit the monadic world once the
constraint has been satisfied. Because of the peculiar nature of the
lock-monad, which set-theoretically corresponds to taking the
\emph{singleton} elements of a set, we can use the very unlock
destructor instead of Moggi's $let_T$ destructor
\cite{Moggi-Computationallambda}, normally used in dealing
with monads. This greatly simplifies the equational theory.

In this paper, we establish the full language theory of the Lax
Logical Framework, \LLFP, by \emph{reducing} it to that of \LF\
itself, \ie, by means of a \emph{metacircular} interpretation of
\LLFP-derivations as \LF\ derivations. This encoding is adequate and
\emph{shallow} enough so that we can transfer to \LLFP\ all the main
properties of \LF. This approach generalizes to \emph{derivations} the
idea underpinning the mapping normally used in the literature to prove
normalization of terms in \LF-like systems \cite{HHP-92,Bar-92}.

Differently from earlier systems with locked types, \eg, \LFP, the system \LLFP\ allows one to reason ``under locks''. This allows for natural encodings of side conditions as appear for instance in the $\xi_v$ rule of the call-by-value $\lambda_v$-calculus, see Section \ref{cbv-sec}.

We discuss encodings in \LLFP\ of various logical systems, thereby
showing that \LLFP\ is the appropriate metalanguage for dealing
formally with side-conditions, as well as external and non-apodictic
evidence. These examples illustrate the extra expressiveness \wrt\
previous systems given by the monadic understanding of locks, namely
the possibility of using \emph{guarded} unlocks
$\Unlock{\P}{M}{\sigma}{\cdot}$, even if the property has not been yet
established. Thus, signatures become much more flexible, hence
achieving the full modularity that we have been looking for in recent
years.  We briefly discuss also a famous system introduced by Fitch
\cite{fitch} of a consistent \emph{Naive Set Theory}.

\medskip
In conclusion, in this paper:
\begin{enumerate}
\item we extend the well understood principle of the \emph{\LF\
    paradigm} for explaining a logic, \ie\ \emph{judgments as types},
  \emph{rules} or \emph{hypothetical judgements as higher-order
    types}, \emph{schemata as higher-order functions}, and
  \emph{quantified variables as bound metalanguage variables}, with
  the new clauses: {\emph{side conditions as monads} and
    \emph{external evidence as monads}};
\item we support the capacity of combining logical systems and
  relating them to software tools using a simple communication
  paradigm via ``wrappers''.
\end{enumerate}

\subsection{Related work}\label{sec:relwork}
This paper builds on the earlier work of the authors
\cite{HLL06,HLLS08,honsell:hal-00906391,Honsell:2013:YFP:2503887.2503896}
and was inspired by the very extensive work on Logical Frameworks by
\cite{pfenning1999system,watkins-02,LF-modulo,NPP05:CMTT,Pientka08:DependentBeluga,belugasys}. The
term \emph{``Lax''} is borrowed from
\cite{fairtlough1997propositional,mendler1991constrained}, and indeed
our system can be viewed as a generalization, to a family of dependent
lax operators, of the work carried out there, as well as Moggi's
\emph{partial} $\lambda$-calculus \cite{moggi1988partial}. A
correspondence between lax modalities and monads in functional
programming was pointed out in \cite{alechina2001,garg2008indexed}.
The connection between constraints and monads in logic programming was
considered in the past, \eg, in
~\cite{NPP05:CMTT,Fairtlough97first-orderlax,fairtlough2001abstraction},
but to our knowledge, this is the first paper which clearly
establishes the correspondence between side conditions and monads in a
\emph{higher-order dependent type theory} and in logical frameworks.

In~\cite{NPP05:CMTT}, the authors introduce a contextual modal logic,
where the notion of context is rendered by means of monadic
constructs. There are points of contact with our work which should be
explored. Here, we only point out that also in their approach they
could have done away with the {\tt let} construct in favour of a
deeper substitution as we have done.

Schr\"oder Heister has discussed in a number of papers, see \eg\
\cite{schroeder2012proof,schroeder2012honour}, various restrictions
and side conditions on rules and on the nature of assumptions that one
can add to logical systems to prevent the arising of paradoxes. There
are some connections between his work and ours and it would be
interesting to compare the bearing of his requirements on side
conditions being ``closed under substitution'' to our notion of
\emph{well-behaved} predicate. Similarly, there are commonalities
between his distinction between \emph{specific} and \emph{unspecific}
variables, and our treatment of free variables in well-behaved
predicates.

\subsection{Some methodological and philosophical remarks on
  \emph{non-apodictic} evidence and formalization}
By the term \emph{non-apodictic evidence} we denote the kind of evidence
which is not derived within the formal system itself. This is the kind
of evidence which normally justifies assumptions or axioms. Often, it
finds its roots in the heuristics which originally inspire the
argument.  Many heuristics are derived from \emph{Physics} or
\emph{analogy}.  Archimedes was a champion of the former, as it is well
documented in his Organon \cite{acerbi}, where he anticipates integral
calculus by conceiving a geometrical figure as composed of thin slices
of a physical object hanging on a balance scale and subject to
gravity. Rather than developing \emph{mathematical physics}, he is, in
fact, performing \emph{physical mathematics}.

Arguments by authority have never been allowed, but the beauty of some
one-line proofs, or of some proofs-without-words, like the jig-saw
puzzle proofs of Pythagoras Theorem, lies precisely in the capacity
that these justifications have of conveying the intuition of why the
statement is plausible. Schopenhauer's \cite{schopenhauer}(ch.15)
criticism of Euclid's ``brilliant abstract nonsense'' proof of
Pythagoras Theorem goes precisely in the direction of defending
\emph{intuitive evidence}.  In order to have a feel for the kind of
evidence we term as non-apodictic, consider the following problem:
given a point inside a convex polyhedron, there exists a face of the
polyhedron such that the projection of the point onto the plane of
that face lies inside the face. How can you formalize adequately the
following non-apodictic argument: such a face has to exist otherwise
we would have a \emph{perpetuum mobile}?

The approach that we put forward in this paper for handling non-apodictic
evidence is simple, but not at all simplistic, given the
fact that the quest for absolute justification leads to an
\emph{infinite regress}. The very \emph{adequacy} of a given
formalization rests ultimately on unformalizable justifications and
even the very \emph{execution} of a rule relies on some \emph{external
  unformalizable} convention, which is manifested only when the rule
is put into practice. As Alain Badiou puts it in \cite{badiou}:
\emph{``ce qui identifie la philosophie ce ne sont pas les r\`{e}gles
  d'un discourse, mais la singularit\'{e} d'un acte''}. The inevitable
infinite regress is captured by the \emph{M\"unchausen
  trilemma}~\cite{albert1968} or by the story of \emph{Achilles and
  the Tortoise} narrated by Lewis Carroll~\cite{carr}\footnote{Notice
  that Girard in The Blind Spot \cite{girard2011blind} provides a
  possibly different appraisal of the same story.}. Ultimately, we can
only ``Just do it!''.

The irreducible and ineliminable role of \emph{conventions} in human
activities, even the apparently most formalizable, has been the object
of interest of many philosophers in the XXth century, \eg\
Wittgenstein or Heidegger. We believe that the first one to point this out
was the italian political philosopher Antonio Gramsci, who wrote in his
Prison Notebooks, 323-43 (Q1112), 1932 ``In acquiring one's conception
of the world, one always belongs to a particular grouping, which is that
of all the social elements that share the same mode of thinking and
acting. We are all conformists of some conformism or other, always
man-in-the-mass or collective man. The question is this: of what
historical type is the conformism, the mass humanity to which one
belongs?''

Different proof tools, or proof search mechanisms are simply other kinds
of conformisms.  Summing up, \LLFP\ makes it possible to invoke our
conformism within a Logical Framework, and it
is formally rigorous in keeping track of when we do that and in
permitting us to explain it away when we can.

\section{The system \LLFP}\label{sec:lfp}
In this section, following the standard pattern and conventions of
\cite{HHP-92}, we introduce the syntax and the rules of \LLFP: in
Figure~\ref{fig:lfpsyntax}, we give the syntactic categories of \LLFP,
namely signatures, contexts, kinds, families (\ie, types) and objects
(\ie, terms), while the main one-step $\beta\L$-reduction rules appear
in Figure \ref{fig:mainred}.

The rules for one-step closure under context for kinds are presented
in~\prettyref{fig:cuckinds}, while those for families and objects are
presented in~\prettyref{fig:blcc}, and~\prettyref{fig:cucobjects}. We
denote the reflexive and transitive closure of $\evalBL$ by
$\multievalBL$. Hence, $\beta\L$-definitional equality is defined in
the standard way, as the reflexive, symmetric, and transitive closure
of $\beta\L$-reduction on kinds, families, and objects, as illustrated
in Figure~\ref{fig:bldefeq}. The language of \LLFP\ is the same as
that of \LFP\ \cite{honsell:hal-00906391}. In particular, \wrt\
classical \LF, we add the \emph{lock-types} constructor ($\mathcal L$)
for building types of the shape $\Lock{\P}{N}{\sigma}{\rho}$, where
$\P$ is a predicate on typed judgements. Correspondingly, at the object
level, we introduce the \emph{constructor} lock ($\mathcal L$) and the
\emph{destructor} unlock ($\mathcal U$).  The intended meaning of the
$\Lock{\P}{N}{\sigma}{\cdot}$ constructors is that of \emph{logical
  filters}. Locks can be viewed also as a generalization of the
\emph{Lax} modality of
\cite{fairtlough1997propositional,mendler1991constrained}.  One of the
points of this paper is to show that they can be viewed also as
\emph{monads}.
\begin{figure}[t!]
\begin{normalsize}
  \begin{center}
      $
      \begin{array}{rcl@{\quad}rcl}
        \Sigma & \in & \emph{Signatures} &
        \Sigma & ::= & \emptyset \mid \Sigma,a \of K \mid \Sigma, c \of \sigma
        \\[1mm]

        \hideshow{
          \Gamma & \in & \emph{Contexts} &
          \Gamma & ::= & \emptyset \mid \Gamma, x \of \sigma
          \\[1mm]
        }

        K & \in &  \emph{Kinds}  & K & ::= &
        \Type \mid {\Prod x \sigma K}
        \\[1mm]

        \sigma,\tau, \rho & \in & \emph{Families (Types)}
        & \sigma & ::= & a \mid {\Prod x \sigma \tau} \mid {\App \sigma N} \mid
        {\Lock {\P} N \sigma \rho}
        \\[1mm]

        M, N & \in & \emph{Objects} &
        M & ::= &
        c \mid x \mid {\Abs x \sigma M} \mid {\App M N} \mid
        \Lock {\P} N \sigma M \mid {\Unlock {\P} N \sigma M}
      \end{array}
      $
    \end{center}
  \caption{The pseudo-syntax of \LLFP}
  \label{fig:lfpsyntax}
\end{normalsize}
\end{figure}

\begin{figure}[t!]
\begin{normalsize}
\begin{center}
  $({\Abs x \sigma M}) \at N \evalBL M[N/x]$ $\BOMain$\qquad
  $\Unlock {\P} N \sigma {\Lock {\P} N \sigma M} \evalBL M$
  $\LOMain$
  \caption{Main one-step-$\beta\mathcal{L}$-reduction rules}
  \label{fig:mainred}
\end{center}
\end{normalsize}
\end{figure}

\begin{figure}
  \begin{center}
    \begin{tabularx}{16cm}{m{5cm}m{2cm}m{5cm}m{2cm}}
      \begin{prooftree}
        \AxiomC{$\sigma \evalBL \sigma'$}
        \UnaryInfC{${\Prod x \sigma \tau} \evalBL {\Prod x {\sigma'} \tau}$}
      \end{prooftree} & $\CCFPa$ &
      \begin{prooftree}
        \AxiomC{$\tau \evalBL \tau'$}
        \UnaryInfC{${\Prod x \sigma \tau} \evalBL {\Prod x \sigma {\tau'}}$}
      \end{prooftree} & $\CCFPb$ \\
      \begin{prooftree}
        \AxiomC{$\sigma \evalBL \sigma'$}
        \UnaryInfC{${\App \sigma N} \evalBL {\App {\sigma'} N}$}
      \end{prooftree} & $\CCFAa$ &
      \begin{prooftree}
        \AxiomC{$N \evalBL N'$}
        \UnaryInfC{${\App \sigma N} \evalBL {\App \sigma {N'}}$}
      \end{prooftree} & $\CCFAb$ \\
      \begin{prooftree}
        \AxiomC{$N \evalBL N'$}
        \UnaryInfC{${\Lock {\P} N \sigma \rho} \evalBL {\Lock {\P} {N'} \sigma \rho}$}
      \end{prooftree} & $\CCFLa$ &
      \begin{prooftree}
        \AxiomC{$\sigma \evalBL \sigma'$}
        \UnaryInfC{${\Lock {\P} N \sigma \rho} \evalBL {\Lock {\P} N {\sigma'} \rho}$}
      \end{prooftree} & $\CCFLb$ \\
      \begin{prooftree}
        \AxiomC{$\rho \evalBL \rho'$}
        \UnaryInfC{${\Lock {\P} N \sigma \rho} \evalBL {\Lock {\P} N \sigma {\rho'}}$}
      \end{prooftree} & $\CCFLc$ & & \\
    \end{tabularx}
  \end{center}
  \caption{$\beta\mathcal{L}$-closure-under-context for families}
  \label{fig:blcc}
\end{figure}

\begin{figure}[!h]
	\begin{center}
		\begin{tabularx}{16cm}{m{5cm}m{2cm}m{5cm}m{2cm}}
			\begin{prooftree}
				\AxiomC{$\sigma \evalBL \sigma'$}
				\UnaryInfC{${\Prod x \sigma K} \evalBL {\Prod x {\sigma'} K}$}
			\end{prooftree} & $\CCKPa$ &
			\begin{prooftree}
				\AxiomC{$K \evalBL K'$}
				\UnaryInfC{${\Prod x \sigma K} \evalBL {\Prod x \sigma {K'}}$}
			\end{prooftree} & $\CCKPb$ \\
		\end{tabularx}
  \end{center}
  \caption{$\beta\mathcal{L}$-closure-under-context for kinds}
  \label{fig:cuckinds}
\end{figure}

\begin{figure}[!h]
	\begin{center}
		\begin{tabularx}{16cm}{m{5cm}m{2cm}m{5cm}m{2cm}}
			\begin{prooftree}
      	\AxiomC{$\sigma \evalBL \sigma'$}
      	\UnaryInfC{${\Abs x \sigma M} \evalBL {\Abs x {\sigma'} M}$}
			\end{prooftree} & $\CCOAba$ &
        \begin{prooftree}
 		     \AxiomC{$M \evalBL M'$}
 		     \UnaryInfC{${\Abs x \sigma M} \evalBL {\Abs x \sigma {M'}}$}
			\end{prooftree} & $\CCOAbb$ \\
			\begin{prooftree}
  	   		 \AxiomC{$M \evalBL M'$}
  			    \UnaryInfC{${\App M N} \evalBL {\App {M'} N}$}
			\end{prooftree} & $\CCOApa$ &
			\begin{prooftree}
  		    \AxiomC{$N \evalBL N'$}
 		    \UnaryInfC{${\App M N} \evalBL {\App M {N'}}$}
			\end{prooftree} & $\CCOApb$ \\
			\begin{prooftree}
      		\AxiomC{$N \evalBL N'$}
      		\UnaryInfC{${\Lock {\P} N \sigma M} \evalBL {\Lock {\P} {N'} \sigma M}$}
			\end{prooftree} & $\CCOLa$ &
			\begin{prooftree}
   		   \AxiomC{$\sigma \evalBL \sigma'$}
   		   \UnaryInfC{${\Lock {\P} N \sigma M} \evalBL {\Lock {\P} N {\sigma'} M}$}
			\end{prooftree} & $\CCOLb$ \\
			\begin{prooftree}
      		\AxiomC{$M \evalBL M'$}
      		\UnaryInfC{${\Lock {\P} N \sigma M} \evalBL {\Lock {\P} N \sigma {M'}}$}
			\end{prooftree} & $\CCOLc$ &
			\begin{prooftree}
      		\AxiomC{$N \evalBL N'$}
      		\UnaryInfC{${\Unlock {\P} N \sigma M} \evalBL {\Unlock {\P} {N'} \sigma M}$}
			\end{prooftree} & $\CCOUa$ \\
			\begin{prooftree}
   		   \AxiomC{$\sigma \evalBL \sigma'$}
   		   \UnaryInfC{${\Unlock {\P} N \sigma M} \evalBL {\Unlock {\P} N {\sigma'} M}$}
			\end{prooftree} & $\CCOUb$ &
			\begin{prooftree}
      		\AxiomC{$M \evalBL M'$}
      		\UnaryInfC{${\Unlock {\P} N \sigma M} \evalBL {\Unlock {\P} N \sigma {M'}}$}
			\end{prooftree} & $\CCOUc$ \\
		\end{tabularx}
  \end{center}
  \caption{$\beta\mathcal{L}$-closure-under-context for objects}
  \label{fig:cucobjects}
\end{figure}

\begin{figure}[!h]
  \begin{center}
    \begin{tabularx}{16cm}{m{4cm}m{2cm}m{5cm}m{2cm}}
			\begin{prooftree}
      	\AxiomC{$T \evalBL T'$}
      	\UnaryInfC{$T \eqBL T'$}
			\end{prooftree} & $\BLEqMain$ &
			\begin{prooftree}
 		     \AxiomC{}
 		     \UnaryInfC{$T \eqBL T$}
			\end{prooftree} & $\BLEqRefl$ \\
			\begin{prooftree}
  	   		 \AxiomC{$T \eqBL T'$}
  			    \UnaryInfC{$T' \eqBL T$}
			\end{prooftree} & $\BLEqSym$ &
			\begin{prooftree}
  		    \AxiomC{$T  \eqBL T'$}
  		    \AxiomC{$T' \eqBL T''$}
 		    \BinaryInfC{$T \eqBL T''$}
			\end{prooftree} & $\BLEqTrans$ \\
    \end{tabularx}
  \end{center}
  \caption{$\beta\L$-definitional equality}
  \label{fig:bldefeq}
\end{figure}
For the sake of generality,
we allow declarations of the form
$x\of\Lock{\P}{N}{\sigma}{\tau}$ in contexts, \ie, we allow one to declare
variables ranging over lock-types, albeit this is not used in
practice.

\begin{figure}[t!]
  \begin{normalsize}
    \begin{center}
      $
      \begin{array}[t]{l}

        \mbox{\sf Signature~rules}\hfill
        \\[2mm]

        \infer[(S{\cdot}Empty)]
        {\emptyset\ \sig}
        {}
        \\[3mm]

        \infer[(S{\cdot}Kind)]
        {\Sigma, a \of K\ \sig}
        {\begin{array}{l@{\quad}l}
            \VDASHS K & a \not \in \Dom(\Sigma)
          \end{array}}
        \\[3mm]

        \infer[(S{\cdot}T\!ype)]
        {\Sigma, c \of \sigma\ \sig}
        {\begin{array}{l@{\quad}l}
            \VDASHS \sigma \of \Type & c \not \in \Dom(\Sigma)
          \end{array}}
        \\[3mm]

        \mbox{\sf Context~rules}\hfill
        \\[2mm]

        \infer[(C{\cdot}Empty)]
        {\VDASHS \emptyset}
        {\Sigma\ \sig}
        \\[3mm]

        \infer[(C{\cdot}T\!ype)]
        {\VDASHS \Gamma, x \of \sigma}
        {\begin{array}{l@{\quad}l}
            \Gamma \VDASHS \sigma \of \Type & x \not \in \Dom(\Gamma)
          \end{array}}
        \\[3mm]

        \mbox{\sf Kind~rules}\hfill
        \\[2mm]

        \infer[(K{\cdot}T\!ype)]
        {\Gamma \VDASHS \Type}
        {\VDASHS \Gamma}
        \\[3mm]

        \infer[(K{\cdot}Pi)]
        {\Gamma \VDASHS {\Prod x \sigma K}}
        {\Gamma, x \of \sigma \VDASHS K}
        \\[4mm]

        \mbox{\sf Family~rules}\hfill
        \\[2mm]

        \infer[(F{\cdot}Const)]
        {\Gamma \VDASHS a : K}
        {\VDASHS \Gamma & a \of K \in \Sigma}
        \\[2mm]

        \infer[(F{\cdot}Pi)]
        {\Gamma \VDASHS {\Prod x \sigma \tau} : \Type}
        {\Gamma, x \of \sigma \VDASHS \tau : \Type}
        \\[-6mm]

      \end{array}
      \rew{67}
      \begin{array}[t]{r}
        \\[1mm]

        \infer[(F{\cdot}App)]
        {\Gamma \VDASHS {\App \sigma N} : K[N/x]}
        {\Gamma \VDASHS \sigma : {\Prod x \tau K} & \Gamma \VDASHS N : \tau}
        \\[3mm]

        \infer[(F{\cdot}Lock)]
        {\Gamma \VDASHS {\Lock {\P} N \sigma {\rho}} : \Type}
        {\Gamma \VDASHS \rho : \Type & \Gamma \VDASHS N : \sigma}
        \\[3mm]

        \infer[(F{\cdot}Conv)]
        {\Gamma \VDASHS \sigma : K'}
        {\Gamma \VDASHS \sigma : K &
          \Gamma \VDASHS K' & K \eqBL K'}
        \\[3mm]

        \mbox{\sf Object~rules}\hfill
        \\[2mm]

        \infer[(O{\cdot}Const)]
        {\Gamma \VDASHS c : \sigma}
        {\VDASHS \Gamma & c \of \sigma \in \Sigma}
        \\[3mm]

        \infer[(O{\cdot}V\!ar)]
        {\Gamma \VDASHS x : \sigma}
        {\VDASHS \Gamma & x \of \sigma \in \Gamma}
        \\[3mm]

        \infer[(O{\cdot}Abs)]
        {\Gamma \VDASHS {\Abs x \sigma M} : {\Prod x \sigma \tau}}
        {\Gamma, x \of \sigma \VDASHS M : \tau}
        \\[3mm]

        \infer[(O{\cdot}App)]
        {\Gamma \VDASHS M \at N : \tau[N/x]}
        {\Gamma \VDASHS M : {\Prod x \sigma \tau} & \Gamma \VDASHS N : \sigma}
        \\[3mm]

        \infer[(O{\cdot}Conv)]
        {\Gamma \VDASHS M : \tau}
        {
          \Gamma \VDASHS M : \sigma & 
          \Gamma \VDASHS \tau : \Type & \sigma \eqBL \tau
        }
        \\[3mm]

        \infer[(O{\cdot}Lock)]
        {\Gamma \VDASHS {\Lock {\P} N \sigma M} : {\Lock {\P} N \sigma {\rho}}}
        {\Gamma \VDASHS M : \rho & \Gamma \VDASHS N : \sigma}
        \\[3mm]

        \infer[(O{\cdot}Top{\cdot}Unlock)]
        {\Gamma \VDASHS {\Unlock {\P} N \sigma M} : \rho}
        {\Gamma \VDASHS M : {\Lock {\P} N \sigma \rho}  &
          \P(\Gamma \VDASHS N : \sigma)}
        \\[3mm]

            \!\!\infer[(F{\cdot}Guarded{\cdot}Unlock)]
            {\Gamma  \VDASHS \Lock {\P} S \sigma { \rho[\Unlock {\P} {S'} {\sigma'} N /x]} : \Type}
            {\Gamma, x\of\tau \VDASHS \Lock {\P} S \sigma {\rho} : \Type & \Gamma \VDASHS N :  \Lock {\P} {S'} {\sigma'} \tau & \sigma \eqBL \sigma' & S \eqBL S' } \qquad\quad
            \\[3mm]

            \infer[(O{\cdot}Guarded{\cdot}Unlock)]
            {\Gamma  \VDASHS {\Lock {\P} S \sigma {M[\Unlock {\P} {S'}
            {\sigma'}{N} /x]}} :
            \Lock {\P} S \sigma { \rho[\Unlock {\P} {S'} {\sigma'} {N} /x]}}
            {\Gamma, x\of\tau \VDASHS \Lock {\P} S \sigma {M} : \Lock {\P} S \sigma {\rho} &
                  \Gamma \VDASHS N :  \Lock {\P} {S'} {\sigma'} {\tau} &
                                                                         \sigma \eqBL \sigma' & S \eqBL S' }
      \end{array}
      $
    \end{center}
  \end{normalsize}
  \caption{The \LLFP\ Type System}
  \label{fig:lfptypesys}
\end{figure}
Following the standard specification paradigm of Constructive Type
Theory, we define lock-types using \emph{introduction},
\emph{elimination}, and \emph{equality rules}. Namely, we introduce a
lock-\emph{constructor} for building objects
$\Lock{\P}{N}{\sigma}{M}$
of type $\Lock{\P}{N}{\sigma}{\rho}$,
via the \emph{introduction rule} $\ROL$.
Correspondingly, we introduce an unlock-\emph{destructor}
$\Unlock{\P}{N}{\sigma}{M}$
via the \emph{elimination rule} $(O{\cdot} Guarded{\cdot} Unlock)$.

The introduction rule of lock-types corresponds to the introduction
rule of monads. The correspondence with the elimination rule for
monads is not so immediate because the latter is normally given using
a $let_T$-construct.
The correspondence becomes clear once we realize that
$let_{T_{{\mathcal
      P}(\Gamma \VDASH S : \sigma )}} x=M\ in \
N$ can be safely replaced by $ N[\Unlock{\P}{S}{\sigma}{M}/
x]$ since the
$\Lock{\P}{S}{\sigma}{\cdot}$-monads
satisfy the property $let_{T_\P}\
x=M\ in \ N\rightarrow N $ if $ x \notin FV(N)$, provided
$x$
occurs \emph{guarded} in $N$,
\ie\ within subterms of the appropriate locked-type.

But, since we do not use the traditional $let_T$ construct in
elimination rules, we have to take care of elimination also at the
level of types by means of the rule
\emph{(F$\cdot$Guarded$\cdot$Unlock)}. Moreover both rules
\emph{(F$\cdot$Guarded$\cdot$Unlock)} and
\emph{(O$\cdot$Guarded$\cdot$Unlock)} need to be merged with
\emph{equality} to preserve subject reduction.

These rules give evidence to the understanding of \emph{locks as
  monads}. Indeed, given a predicate $\P$ and
$\Gamma\VDASHS N : \sigma$, the intended monad $(T_\P,\eta,\mu)$ can
be naturally defined on the term model of \LLFP\ viewed as a
category. In particular
$\eta_{\rho} \eqdef \lambda x \of \rho.\Lock{\P}{N}{\sigma}{x} $ and
$\mu_{\rho} \eqdef \lambda x \of
\Lock{\P}{N}{\sigma}{\Lock{\P}{N}{\sigma}{\rho}}.\
\Lock{\P}{N}{\sigma}{\Unlock{\P}{N}{\sigma}{{\Unlock{\P}{N}{\sigma}{x}}}}
$.
Indeed, if $\Gamma,x\of\rho\VDASHS N:\sigma$ is derivable, the term
for $\eta$ can be easily inferred by applying rules $(O{\cdot}Var)$,
$(O{\cdot}Lock)$, and $(O{\cdot}Abs)$ as follows:
$$
\infer{\Gamma\VDASHS \lambda x\of\rho.{\Lock \P N \sigma x}:\Pi
  x\of\rho.{\Lock \P N \sigma \rho}} {\infer{\Gamma,x\of\rho\VDASHS
    \Lock \P N \sigma x:\Lock \P N \sigma \rho}
  {\Gamma,x\of\rho\VDASHS x:\rho & \Gamma,x\of\rho\VDASHS N:\sigma} }
$$
As for the term for $\mu$, if $\Gamma\VDASHS N:\sigma$ is derivable,
applying weakening and the rules $(O{\cdot}Var)$, $(O{\cdot}Lock)$,
and $(O{\cdot}Guarded{\cdot}Unlock)$, we can derive the following:
$$
\infer{\Gamma,z_2 \of {\Lock \P N \sigma \tau}\VDASHS {\Lock \P N
    \sigma {\Unlock \P N \sigma {z_2}}}:{\Lock \P N \sigma \tau}} {
  \infer{\Gamma,z_2 \of {\Lock \P N \sigma \tau},z_1\of \tau\VDASHS
    {\Lock \P N \sigma {z_1}}:{\Lock \P N \sigma \tau}} {\Gamma,z_2
    \of {\Lock \P N \sigma \tau},z_1\of \tau\VDASHS z_1:\tau &
    \Gamma,z_2\of {\Lock \P N \sigma \tau},z_1\of \tau\VDASHS
    N:\sigma} & \Gamma,z_2\of {\Lock \P N \sigma \tau}\VDASHS z_2 :
  {\Lock \P N \sigma \tau} }
$$
Whence, if
$x\of {\Lock \P N \sigma {\Lock \P N \sigma \tau}}\in\Gamma$, we can
derive the following, applying again rules $(O{\cdot}Var)$, and
$(O{\cdot}Guarded{\cdot}Unlock)$:
$$
\infer{\Gamma\VDASHS {\Lock \P N \sigma {\Unlock \P N \sigma {\Unlock
        \P N \sigma x}}}:{\Lock \P N \sigma \tau}}
{\Gamma,z_2\of{\Lock \P N \sigma \tau}\VDASHS {\Lock \P N \sigma
    {\Unlock \P N \sigma {z_2}}}:{\Lock \P N \sigma \tau} &
  \Gamma,z_2\of {\Lock \P N \sigma \tau}\VDASHS x:{\Lock \P N \sigma
    {\Lock \P N \sigma \tau}}}
$$
And, finally, applying rule $(O{\cdot}Abs)$, we get the term
$\lambda x\of{\Lock \P N \sigma {\Lock \P N \sigma \tau}}.{\Lock \P N
  \sigma {\Unlock \P N \sigma {\Unlock \P N \sigma x}}}$.
Finally, to provide the intended meaning of
$\Lock{\P}{N}{\sigma}{\cdot}$, we need to introduce in \LLFP\ also the
rule $(O{\cdot}Top{\cdot} Unlock)$, which allows for the elimination
of the lock-type constructor if the predicate $\P$ is verified,
possibly \emph{externally}, on an appropriate and derivable
judgement. Figure~\ref{fig:lfptypesys} shows the full typing system of
\LLFP. All  \emph{type equality rules} of \LLFP\ use a notion of
conversion which is a combination of
standard $\beta$-reduction, $\BOMain$, with another notion of
reduction $\LOMain$, called $\L$-reduction. The latter behaves as a
lock-releasing mechanism, erasing the $\U$-$\L$ pair in a term of the
form $\Unlock {\P} N \sigma {\Lock {\P} N \sigma M}$.
Lock-types have been discussed by the authors in a series of papers
\cite{BCKL-03,HLL06,HLLS08,honsell:hal-00906391,Honsell:2013:YFP:2503887.2503896},
but \emph{Guarded Unlock} rules, first suggested
in~\cite{llfp-mfcs2014} have not been fully discussed before. These
rules are crucial, because otherwise in order to release a locked term it is
necessary to query the external oracle \emph{explicitly}, by means of
the rule $(O{\cdot}Top{\cdot}Unlock)$, and obtain a positive answer. This
is rather heavy from the practical point of view, because it might
force the invocation of an external tool more than once for the same
property. Moreover, such properties are not essential to the main
thrust of the proof and one would like to be free to proceed with the
main argument, postponing the verification of ``details'' as much as
possible. But, more importantly, such rules allow us to exploit
hypothetic-general locked judgements in encoding rules, as in the case
of the call-by-value $\lambda$-calculus, see \cite{AHMP-92}, and refer to
terms in locked types by pattern matching.  The improvement in all the
case studies is neat \wrt\ plain old
\LFP~\cite{honsell:hal-00906391}. Namely, even if at a given stage of
the proof development we assume (or are not able, or we do not want to
waste time to verify) a side-condition, we can \emph{postpone} such a
task, by unlocking immediately the given term and by proceeding with
the proof. The lock-type of the term into which we release the
unlocked term will keep track that the verification has to be carried
out, sooner or later.

%
%

The Guarded Unlock rules, namely $(O{\cdot}Guarded{\cdot}Unlock)$ and $(F{\cdot}Guarded{\cdot}Unlock)$ are the novelty w.r.t.~the extended abstract of the present paper which appeared in~\cite{llfp-mfcs2014}. First of all in~\cite{llfp-mfcs2014} there was no Guarded Unlock rule at the level of Type Families, but this appears to be necessary to recover a standard proof of the sub-derivation property. As far as the Guarded Unlock rule at the level of Objects, the new $(O{\cdot}Guarded{\cdot}Unlock)$-rule is, first of all, a restriction of the one
in~\cite{llfp-mfcs2014}. Namely, we require that the subject of the
first premise has an \emph{explicit} outermost lock, otherwise we
can derive unlocked terms also at the top level, if locked variables
appear in the assumptions.
This external lock forces the establishment of all pending constraints before
the nested unlock can surface. We could have ruled out locked assumptions, but this restriction allows for a smoother formulation of  the language theory of \LLFP, as
will be shown in Section \ref{sec:metath}.
Furthermore, the new version of the $(O{\cdot}Guarded{\cdot}Unlock)$-rule uses type equality judgements explicitly. Namely the two minor premises ($\sigma\eqBL\sigma'$ and $S\eqBL S'$) in the $(O{\cdot}Guarded{\cdot}Unlock)$-rule  allow for $\beta\L$-conversion in the subscripts $\sigma$ and $S$ of the lock/unlock operators. This appears to be necessary for subject reduction.

We conclude this section by recalling that, since external predicates
affect reductions in \LLFP, they must be \emph{well-behaved} in order
to preserve subject reduction. And this property is needed for
\emph{decidability}, \emph{relative to} an oracle, which is essential
in \LF 's. Let $\alpha$ be a shorthand for any ``subject of type
predicate'', we introduce the crucial definition:

\begin{defi}[Well-behaved predicates, \cite{honsell:hal-00906391}]
  \label{def:wbred}
  A finite set of predicates $\{ \P_i\}_{i\in I}$ is
  \emph{well-behaved} if each $\P$ in the set satisfies the following
  conditions:
  \begin{enumerate}

  \item {\bf \emph{Closure under signature and context weakening and
        permutation:}}
    \begin{enumerate}
    \item If $\Sigma$ and $\Omega$ are valid signatures such that
      $\Sigma \subseteq \Omega$ and $\P(\Gamma \VDASHS
      \alpha)$, 
      then $\P(\Gamma \VDASHO \alpha)$. 
    \item If $\Gamma$ and $\Delta$ are valid contexts such that
      $\Gamma\subseteq \Delta$ and $\P(\Gamma \VDASHS
      \alpha)$, 
      then \mbox{$\P(\Delta \VDASHS \alpha)$.} 
    \end{enumerate}

  \item{\bf \emph{Closure under substitution:}} If
    $\P(\Gamma, x \of \sigma', \Gamma' \VDASHS N : \sigma)$ and
    $\Gamma \VDASHS N' : \sigma'$, 
    then $\P(\Gamma, \Gamma'[N'/x] \VDASHS N [N'/x] : \sigma[N'/x])$.

  \item{\bf\emph {Closure under reduction:}}
    \begin{enumerate}
    \item If $\P(\Gamma \VDASHS N : \sigma)$ and $N \evalBL N'$,
      then $\P(\Gamma \VDASHS N' : \sigma)$. 
    \item If $\P(\Gamma \VDASHS N : \sigma)$ and
      $\sigma \evalBL \sigma'$, 
      then $\P(\Gamma \VDASHS N :\sigma')$. 
    \end{enumerate}
  \end{enumerate}
\end{defi}

\section{Encoding \LLFP\ in \LF}\label{sec:encodingLLFP}\label{sec:LLFP_in_LF}
In this section we define a very \emph{shallow encoding} of \LLFP\ in
Edinburgh \LF\ \cite{HHP-92}. This translation has two purposes. On
one hand we \emph{explain} the ``gist'' of \LLFP, using the
\emph{normative} \LF\ paradigm. On the other hand, we provide a tool
for transferring properties such as \emph{confluence, normalization}
and \emph{subject reduction} from \LF\ to \LLFP. This approach
generalizes the proof technique used in the literature for proving
normalization of dependent type systems relative to their
corresponding purely propositional variant, \eg, \LF\ relative to the
simply typed $\lambda$-calculus, or the Calculus of Constructions
relative to $\sf F_\omega$ \cite{HHP-92,Bar-92}.

The embedding of \LLFP\ into \LF\ is given by an inductive, \ie\
\emph{compositional}, function which maps derivations in \LLFP\ to
derivations in \LF.  The critical instances occur in relation to
lock-types, as was to be expected. The key idea of the encoding is
based on the analogy \emph{locks as abstractions} and \emph{unlocks as
  applications}. To this end we introduce new type-constants in \LF\ to
represent lock-types in \LLFP, appropriate object constants to
represent external evidence, and use appropriate object variables to
represent hypothetical external evidence. Hence locked types become
$\Pi$-types over such new types and locked terms become abstractions
over such new types.

Before entering into the intricacies of the encoding, we illustrate,
suggestively, how the translation of the basic lock-related rules
would appear in a non-dependent purely propositional fragment, if
there were just one single predicate represented by the proposition,
\ie\ type, $L$:

$$
\begin{array}{c}
  \infer[(O{\cdot}Lock)]{\Gamma \vdash \lambda x\of L. M : L \rightarrow
  A}{\Gamma \vdash M : A}\\[3mm]

  \infer[(O{\cdot}Top{\cdot}Unlock)]{\Gamma \vdash M c : A}{\Gamma \vdash M : L \rightarrow A & \Gamma
                                         \vdash c : L }\\[3mm]

  \infer[(O{\cdot}Guarded{\cdot}Unlock)]{\Gamma \vdash \lambda y\of L.M[N y/x]
  : L \rightarrow A}{\Gamma, x\of B \vdash \lambda y\of L.M: L
  \rightarrow A & \Gamma \vdash N :  L \rightarrow B }
\end{array}
$$

\noindent Resuming full generality, each use of the predicate $\P$
in an \LLFP\ derivation, relative to a context $\Gamma$,
term $N$
and type $\sigma$
is encoded by a corresponding \LF-type denoted by
$\P(x_1,\ldots,x_n,\sigma',N')$
where $\{x_1,\ldots,x_n\}\equiv
\Dom(\Gamma)$\footnote{By inspection on the clauses of the encoding function
  $\E{\Sigma}{\Gamma}$ (introduced later in this section),
  it is clear that, if $\{x_1,\ldots,x_n\}$
  is the domain of the original typing context in \LLFP, then it will
  also be the domain of its encoding in \LF.} and $\sigma'$,
$N'$
are the encodings in \LF\ of $\sigma$
and $N$,
respectively. However, since \LF\ is not a polymorphic type theory, we
cannot feed $\sigma'$
directly to the constant $\P$.
Hence, we use a simple ``trick'' representing $\sigma'$
indirectly by means of the identity function $\lambda
x\of\sigma'.x$ (or
$I_{\sigma'}$
for short).  Thus for each predicate $\P$
in \LLFP, we introduce in \LF\ two families of constants depending on
the environment $\Gamma
\equiv x_1\of\sigma_1, \ldots,
x_n\of\sigma_n$, the signature $\Sigma$, and the type $\Gamma \VDASHS
\sigma: \Type$ as follows:
$$
P^{\Sigma}_{\Gamma}:\Pi x_1\of\sigma'_1 \ldots
x_n\of\sigma'_n. (\sigma'\rightarrow
\sigma')\rightarrow \sigma' \rightarrow\Type$$ and
$$ c_{\P^{\Sigma}_{\Gamma}}: \Pi x_1\of \sigma'_1
\ldots x_n\of \sigma'_n.\Pi x\of{\sigma'}\rightarrow
{\sigma'}.y\of{\sigma'}. (\P^{\Sigma}_{\Gamma}x_1
\ldots x_n x y).$$
where $\sigma'_i$ ($1\leq i\leq n$) and $\sigma'$ are the encodings in \LF\ of $\sigma_i$ and $\sigma$, respectively.
The former constants are used to encode the lock-type, in such a way
that the derivation of the term $(c_{\P^\Sigma_{\Gamma}}\
x_1 \ldots x_n\ I_{\sigma'}\ N')$ or of a variable of type $(P^\Sigma_{\Gamma}\
x_1 \ldots x_n\ I_{\sigma'}\ N')$ will encode in \LF\ the fact that the external judgment
$\P(\Gamma\VDASHS
N :
\sigma)$ of \LLFP\ holds or it is assumed to hold. Notice that the
properties of \emph{well-behaved} predicates ensure precisely that
such encodings can be safely introduced without implicitly enforcing
the validity of any spurious judgement. In the following, we will
abbreviate the list $x_1,x_2,\ldots,x_n$
as $\vec{x}$,
whenever it will be clear from the context the origin of the
$x_i$'s.
Moreover, we will drop the $\Sigma$
and $\Gamma$
in the notation of the constants $\P^{\Sigma}_{\Gamma}$
and $c_{\P^{\Sigma}_{\Gamma}}$.

For the above reasons, if the judgment labelling the root of a
derivation tree in \LLFP\ is, $\Gamma\VDASHS\ M:\sigma$, the signature
of the corresponding judgement in \LF\ is not, in general, a
one-to-one translation of the declarations contained in
$\Sigma$. Further constants are needed for encoding predicates and
external evidence, be it concrete if it derives from the oracle's call
and a $(Top{\cdot}Unlock)$ rule, or hypothetical if it derives from a
$(Guarded{\cdot}Unlock)$ rule.

More precisely, the encoding function, denoted by $\E{}{}$ in the
following, will yield, as the translation progresses, an \LF-signature
which possibly increases from the initially empty one, with a
\begin{enumerate}\sloppy
\item possibly fresh $\P$-like constant whenever a lock or unlock
  operator is introduced in rules $(F{\cdot}Lock)$ and $(O{\cdot}Lock)$ and
  $(O{\cdot}Top{\cdot}Unlock)$ and $(F{\cdot}Guarded{\cdot}Unlock)$, and
  $(O{\cdot}Guarded{\cdot}Unlock)$;
\item possibly fresh $c_{\P}$ constant, witnessing the \emph{external
    evidence}, introduced in rule $(O{\cdot}Top{\cdot}Unlock)$. Notice
  that the translation of the first premise of that rule, which
  involves the lock-type already provides the constant $\P$.
\end{enumerate}

\noindent As a consequence, in translating a rule which has two or more premises it is
necessary to \emph{merge} the resulting signatures from the
corresponding translations. The function \textsf{Merge} concatenates
the declarations in the input signatures passed as arguments, pruning
out possible duplications. Merging signatures requires
\emph{engrafting} subtrees, of the appropriate derivations, in the
derivations of the original signatures, thus establishing the validity
of the ``augmented'' counterparts. We denote this, ultimately
straightforward, ``rearrangement'' with the notation $(\D)^+$ in
Figures~\ref{fig:sig_enc}, \ref{fig:ctx_enc}, \ref{fig:kind_enc},
\ref{fig:family_enc1}, \ref{fig:family_enc2}, \ref{fig:obj_enc1},
\ref{fig:obj_enc2}, \ref{fig:obj_lock_enc1}, \ref{fig:fam_unlock_enc},
and \ref{fig:obj_lock_enc2}.

In the following, for the sake of simplicity and readability, we will
denote the result of the application of the mapping function
$\E{\Sigma}{\Gamma}$ (see Figure~\ref{fig:terms_enc}) on terms with an
overline ($\overline{\phantom{x}}$), whenever it will be clear which
are the signature and the environment involved. The notation is also extended to signatures and typing environments in the obvious way.

Finally we point out that the function $\E{}{}$ induces a
\emph{compositional} map from kinds, families, and objects in \LLFP\
to the corresponding categories in \LF. We denote such a map by
$\ET{\Sigma}{\Gamma}$ and we provide an independent inductive
definition in Figure~\ref{fig:terms_enc}. It receives as input
parameters the signature $\Sigma$ and the typing context $\Gamma$
synthesized by the map $\E{}{}$ encoding derivations.
\begin{figure}
  \caption{Encoding of signature rules}\label{fig:sig_enc}
  \begin{small}
    {\renewcommand{\arraystretch}{2}
      \begin{tabular}{|l|}\hline
        \\[-2em]
        $\E{\Sigma}{\Gamma}\left(\begin{array}{c}\infer[(S{\cdot}Empty)] {\emptyset\ \sig} {-}\end{array}\right)
        \Longrightarrow\begin{array}{c}\infer[] {\emptyset\ \sig} {}\end{array}
        $\hfil\\[3mm]\hline
        \\[-2em]
        $\E{\Sigma}{\Gamma}\left(\begin{array}{c}\infer[(S{\cdot}Kind)] {\Sigma', a \of K\ \sig}
                                               {\infer{\VDASH_{\Sigma'} K}{\D} & a \not \in \Dom(\Sigma')}\end{array} \right)  \Longrightarrow \begin{array}{c}\infer[] {\Sigma'', a \of K'\ \sig}
                                                              {\infer{\VDASH_{\Sigma''} K'}{\D'} & a \not \in \Dom(\Sigma'')}\end{array} $\hfil\\[3mm]
        \hfill where
        $\E{\Sigma''}{\Gamma}\left(\begin{array}{c}\infer{\VDASH_{\Sigma'} K}{\D}\end{array}\right) \Longrightarrow \begin{array}{c}\infer{\VDASH_{\Sigma''}K'}{\D'}\end{array}$\\[3mm]\hline
        \\[-2em]
        $\E{\Sigma}{\Gamma}\left(\begin{array}{c}\infer[(S{\cdot}Type)] {\Sigma', c \of \sigma\ \sig}
                                               {\infer{\VDASH_{\Sigma'} \sigma:\Type}{\D} & c \not \in \Dom(\Sigma')} \end{array}\right) \Longrightarrow \begin{array}{c}\infer[]  {\Sigma'', c \of \sigma'\ \sig}
                                                                {\infer{\VDASH_{\Sigma''} \sigma':\Type}{\D'} & c \not \in \Dom(\Sigma'')}\end{array} $\hfil\\[3mm]
        \hfill where
        $\E{\Sigma''}{\Gamma}\left(\begin{array}{c}\infer{\VDASH_{\Sigma'} \sigma:\Type}{\D}\end{array}\right) \Longrightarrow \begin{array}{c}\infer{\VDASH_{\Sigma''}\sigma':\Type}{\D'}\end{array}$\\[3mm]\hline
  \end{tabular}
}
\end{small}
\end{figure}

\begin{figure}
  \caption{Encoding of typing context rules}\label{fig:ctx_enc}
  \begin{small}
    {\renewcommand{\arraystretch}{2}
      \begin{tabular}{|l|}\hline
        \\[-2em]
        $\E{\Sigma}{\Gamma}\left(\begin{array}{c}\infer[(C{\cdot}Empty)]  {\VDASHS \emptyset}  {\infer{\Sigma'\ \sig}{\D}}\end{array} \right)\Longrightarrow \begin{array}{c}\infer[]  {\VDASH_{\Sigma''}\emptyset}  {\infer{\Sigma''\ \sig}{\D'}}\end{array}  $\hfill
        where $\E{\Sigma}{\Gamma}\left(\begin{array}{c}\infer{\Sigma'\ \sig}{\D}\end{array}\right) \Longrightarrow \begin{array}{c}\infer{\Sigma''\ \sig}{\D'}\end{array}$
        \\[5mm]\hline
        \\[-2em]
        $\E{\Sigma}{\Gamma}\left(\begin{array}{c}\infer[(C{\cdot}Type)] {\VDASH_{\Sigma'}\Gamma', x \of \sigma}  {\infer{\VDASH_{\Sigma'}\Gamma'}{\D_1} & \infer{\Gamma'\VDASH_{\Sigma'} \sigma:\Type}{\D_2} & (1) }\end{array} \right) \Longrightarrow \begin{array}{c}\infer {\VDASH_{\Sigma^{iv}} \Gamma'', x \of \sigma' } {\infer{\VDASH_{\Sigma^{iv}}\Gamma''}{(\D'_1)^+} & \infer{\Gamma''\VDASH_{\Sigma^{iv}} \sigma':\Type}{(\D'_2)^+} & (2)}\end{array} $\hfil\\[3mm]
        \hfill where
        $\E{\Sigma}{\Gamma}\left(\begin{array}{c}\infer{\VDASH_{\Sigma'}\Gamma'}{\D_1}\end{array}\right) \Longrightarrow \begin{array}{c}\infer{\VDASH_{\Sigma''}\Gamma''}{\D'_1}\end{array}$, and $\E{\Sigma}{\Gamma}\left(\begin{array}{c}\infer{\Gamma'\VDASH_{\Sigma'} \sigma:\Type}{\D_2}\end{array}\right)
        \Longrightarrow \begin{array}{c}\infer{\Gamma''\VDASH_{\Sigma'''}\sigma':\Type}{\D'_2}\end{array}$, and \\[3mm]
        \hfill $\Sigma^{iv}\eqdef \M(\Sigma'',\Sigma''')$, and  $(1) \eqdef x \not \in \Dom(\Gamma')$, and $(2) \eqdef x \not \in \Dom(\Gamma'')$\\[3mm]\hline
  \end{tabular}
}
\end{small}
\end{figure}

\begin{figure}
  \caption{Encoding of kind rules}\label{fig:kind_enc}
  \begin{small}
    {\renewcommand{\arraystretch}{1.4}
      \begin{tabular}{|l|}\hline
        \\[-1.5em]
        $\E{\Sigma}{\Gamma}\left(\begin{array}{c}\infer[(K{\cdot}T\!ype)] {\Gamma' \VDASH_{\Sigma'} \Type} {\infer{\VDASH_{\Sigma'} \Gamma'}{\D}}\end{array} \right) \Longrightarrow \begin{array}{c}\infer[]  {\Gamma'' \VDASH_{\Sigma''} \Type}  {\infer{\VDASH_{\Sigma''}\Gamma''}{\D'}}\end{array}
        $\hfill
        \hfill where
        $\E{\Sigma}{\Gamma}\left(\begin{array}{c}\infer{\VDASH_{\Sigma'} \Gamma'}{\D}\end{array}\right) \Longrightarrow \begin{array}{c}\infer{\VDASH_{\Sigma''}\Gamma''}{\D'}\end{array}$\\[5mm]\hline
        \\[-1.5em]
        $\E{\Sigma}{\Gamma}\left(\begin{array}{c}\infer[(K{\cdot}Pi)]  {\Gamma' \VDASH_{\Sigma'} {\Prod x \sigma K}}  {\infer{\Gamma', x \of \sigma \VDASH_{\Sigma'} K}{\D}}\end{array} \right) \Longrightarrow \begin{array}{c}\infer[] {\Gamma''\VDASH_{\Sigma''}\Prod x {\sigma'} {K'}} {\infer{\Gamma'',x\of\sigma'\VDASH_{\Sigma''}{K'}}{\D'}}\end{array} $\hfill\\[3mm]
        \hfill where
        $\E{\Sigma}{\Gamma}\left(\begin{array}{c}\infer{\Gamma', x\of \sigma \VDASH_{\Sigma'} K}{\D}\end{array}\right) \Longrightarrow \begin{array}{c}\infer{\Gamma'',x:\sigma'\VDASH_{\Sigma''}{K'}}{\D'}\end{array}$\\[5mm]
        \hline
  \end{tabular}
}
\end{small}
\end{figure}

\begin{figure}
  \caption{Encoding of family rules - Pt.1}\label{fig:family_enc1}
  \begin{small}
    {\renewcommand{\arraystretch}{1.4}
      \begin{tabular}{|l|}\hline
        \\[-1.5em]
        $\E{\Sigma}{\Gamma}\left(\begin{array}{c}\infer[(F{\cdot}Const)] {\Gamma' \VDASH_{\Sigma'} a : K} {\infer{\VDASH_{\Sigma'} \Gamma'}{\D} & a \of K \in \Sigma'}\end{array} \right)\Longrightarrow \begin{array}{c}\infer[] {\Gamma''\VDASH_{\Sigma''}a:K'}  {\infer{\VDASH_{\Sigma''} \Gamma''}{\D'} & a \of K' \in \Sigma''}\end{array}$\hfill\\[3mm]
        \hfill where
        $\E{\Sigma}{\Gamma}\left(\begin{array}{c}\infer{\VDASH_{\Sigma'} \Gamma'}{\D}\end{array}\right)
        \Longrightarrow \begin{array}{c}\infer{\VDASH_{\Sigma''} \Gamma''}{\D'}\end{array}$\\[5mm]\hline
        \\[-1.5em]
        $\E{\Sigma}{\Gamma}\left(\begin{array}{c}\infer[(F{\cdot}Pi)]  {\Gamma' \VDASH_{\Sigma'} {\Prod x \sigma \tau} : \Type} {\infer{\Gamma', x \of \sigma \VDASH_{\Sigma'} \tau : \Type}{\D}}\end{array} \right)  \Longrightarrow \begin{array}{c}\infer[] {\Gamma'' \VDASH_{\Sigma''} {\Prod x \sigma' \tau'} : \Type} {\infer{\Gamma'', x \of \sigma' \VDASH_{\Sigma''} \tau' : \Type}{\D'}}\end{array}
        $\hfil\\[3mm]
        \hfill where $\E{\Sigma}{\Gamma}\left(\begin{array}{c}\infer{\Gamma', x \of \sigma \VDASH_{\Sigma'} \tau : \Type}{\D}\end{array}\right)
        \Longrightarrow \begin{array}{c}\infer{\Gamma'', x \of \sigma' \VDASH_{\Sigma''} \tau' : \Type}{\D'}\end{array}$\\[3mm]\hline
        \\[-1.5em]
        $\E{\Sigma}{\Gamma}\left(\begin{array}{c}\infer[(F{\cdot}App)] {\Gamma' \VDASH_{\Sigma'} {\App \sigma N} : K[N/x]}  {\infer{\Gamma' \VDASH_{\Sigma'} \sigma : {\Prod x \tau K}}{\D_1} & \infer{\Gamma' \VDASH_{\Sigma'} N : \tau}{\D_2}}\end{array}\right)\Longrightarrow \begin{array}{c}\infer  {\Gamma'' \VDASH_{\Sigma^{iv}} {\App \sigma' N'} : K'[N'/x]}  {\infer{\Gamma'' \VDASH_{\Sigma^{iv}} \sigma' : {\Prod x \tau' K'}}{(\D'_1)^+} & \infer{\Gamma'' \VDASH_{\Sigma^{iv}} N' : \tau'}{(\D'_2)^+}}\end{array}  $\hfil\\[3mm]
        \hfill where $\E{\Sigma}{\Gamma}\left(\begin{array}{c}\infer{\Gamma' \VDASH_{\Sigma'} \sigma : {\Prod x \tau K}}{\D_1}\end{array}\right)
        \Longrightarrow \begin{array}{c}\infer{\Gamma'' \VDASH_{\Sigma''} \sigma' : {\Prod x \tau' K'}}{\D'_1}\end{array}$, and \\[3mm]
        \hfill $\E{\Sigma}{\Gamma}\left(\begin{array}{c}\infer{\Gamma' \VDASH_{\Sigma'} N : \tau}{\D_2}\end{array}\right)
        \Longrightarrow \begin{array}{c}\infer{\Gamma'' \VDASH_{\Sigma'''} N' : \tau'}{\D'_2}\end{array}$,  and $\Sigma^{iv}\eqdef \M(\Sigma'',\Sigma''')$\\\hline
      \end{tabular}
    }
  \end{small}
\end{figure}

\begin{figure}
  \caption{Encoding of family rules - Pt.2}\label{fig:family_enc2}
  \begin{small}
    {\renewcommand{\arraystretch}{1.4}
      \begin{tabular}{|l|}\hline
        \\[-1.5em]
        $\E{\Sigma}{\Gamma}\left(\begin{array}{c}\infer[(\ast)]{\Gamma' \VDASH_{\Sigma'} {\Lock {\P} N \sigma {\rho}} : \Type} {\infer{\Gamma' \VDASH_{\Sigma'} \rho : \Type}{\D_1} & \infer{\Gamma' \VDASH_{\Sigma'} N : \sigma}{\D_2}}\end{array} \right)$ $\Longrightarrow $ $\begin{array}{c}\infer {\Gamma'' \VDASH_{\Sigma^{v}} \Pi y\of (\P\ \vec{x}\ I_{\sigma'}\ N').\rho':\Type} {\infer{\Gamma'' \VDASH_{\Sigma^{v}} \rho' : \Type}{(\D'_1)^+} & \infer{\Gamma'' \VDASH_{\Sigma^{v}} N' : \sigma'}{(\D'_2)^+}}\end{array} $\hfil\\[3mm]
        $\ast \eqdef (F{\cdot}Lock)$ \hfill where $\E{\Sigma}{\Gamma}\left(\begin{array}{c}\infer{\Gamma' \VDASH_{\Sigma'} \rho : \Type}{\D_1}\end{array}\right)  \Longrightarrow\begin{array}{c}\infer{\Gamma'' \VDASH_{\Sigma''} \rho' : \Type}{\D'_1}\end{array}$, and\\[3mm]
        \hfill $\E{\Sigma}{\Gamma}\left(\begin{array}{c}\infer{\Gamma'  \VDASH_{\Sigma'} N : \sigma}{\D_2}\end{array}\right) \Longrightarrow\begin{array}{c}\infer{\Gamma'' \VDASH_{\Sigma'''} N' : \sigma'}{\D'_2}\end{array}$, and \\[3mm]
        \hfill $x_1,\ldots,x_n\eqdef \vec{x} \eqdef \Dom(\Gamma')$, and $\Sigma^{iv}\eqdef \M(\Sigma'',\Sigma''')$, and \\[3mm]
        \hfill $\Sigma^{v}\eqdef \left\{\begin{array}{lr}\Sigma^{iv} & \mbox{ if }\P\in \Dom(\Sigma^{iv})\\\Sigma^{iv},\P\of \Pi x_1\of \sigma_1 \ldots\ x_n\of\sigma_n. (\sigma'\rightarrow \sigma')\rightarrow \sigma' \rightarrow\Type & \mbox{ otherwise} \end{array}\right.$
        \\[5mm]\hline
        \\[-1.5em]
        $\E{\Sigma}{\Gamma}\left(\begin{array}{c}\infer[(\ast)]  {\Gamma' \VDASH_{\Sigma'} \sigma : K'} {\infer{\Gamma' \VDASH_{\Sigma'} \sigma : K}{\D_1} & \infer{\Gamma' \VDASH_{\Sigma'} K'}{\D_2} & \infer{K \eqBL K'}{\D_3}}\end{array} \right) \Longrightarrow \begin{array}{c}\infer{\Gamma'' \VDASH_{\Sigma^{iv}} \sigma' : K'''} {\infer{\Gamma'' \VDASH_{\Sigma^{iv}} \sigma' : K''}{(\D'_1)^+} & \infer{\Gamma'' \VDASH_{\Sigma^{iv}} K'''}{(\D'_2)^+} & \infer{K'' \eqB K'''}{\D'_3}}\end{array} $\hfil\\[3mm]
        $\ast \eqdef (F{\cdot}Conv)$  \hfill where $\E{\Sigma}{\Gamma}\left(\begin{array}{c}\infer{\Gamma' \VDASH_{\Sigma'} \sigma : K}{\D_1}\end{array}\right) \Longrightarrow\begin{array}{c}\infer{\Gamma'' \VDASH_{\Sigma''} \sigma' : K''}{\D'_1}\end{array}$, and \\[3mm]
        \hfill $\E{\Sigma}{\Gamma}\left(\begin{array}{c}\infer{\Gamma'  \VDASH_{\Sigma'} K'}{\D_2}\end{array}\right) \Longrightarrow\begin{array}{c}\infer{\Gamma'' \VDASH_{\Sigma'''} K'''}{\D'_2}\end{array}$, and  \\[3mm]
        \hfill $\Sigma^{iv}\eqdef \M(\Sigma'',\Sigma''')$, and\\[3mm]
        \hfill $\E{\Sigma^{iv}}{\Gamma^{iv}}\left(\begin{array}{c}\infer{K \eqBL K'}{\D_3}\end{array}\right) \Longrightarrow\begin{array}{c}\infer{K'' \eqB K'''}{\D'_3}\end{array}$\\[3mm]\hline
      \end{tabular}
    }
  \end{small}
\end{figure}

\begin{figure}
  \caption{Encoding of the ``standard'' object rules -  Pt.1}\label{fig:obj_enc1}
  \begin{small}
    {\renewcommand{\arraystretch}{1.4}
      \begin{tabular}{|l|}\hline
        \\[-1.5em]
        $\E{\Sigma}{\Gamma}\left(\begin{array}{c}\infer[(O{\cdot}Const)] {\Gamma' \VDASH_{\Sigma'} c : \sigma} {\infer{\VDASH_{\Sigma'} \Gamma'}{\D} & c \of \sigma \in \Sigma'}\end{array} \right) \Longrightarrow\begin{array}{c}\infer[] {\Gamma'' \VDASH_{\Sigma''} c : \sigma'}  {\infer{\VDASH_{\Sigma''} \Gamma''}{\D'} & c \of \sigma' \in \Sigma'' }\end{array} $\hfil where $\E{\Sigma}{\Gamma}\left(\begin{array}{c}\infer{\VDASH_{\Sigma'} \Gamma'}{\D}\end{array}\right)  \Longrightarrow\begin{array}{c}\infer{\VDASH_{\Sigma''} \Gamma''}{\D'}\end{array}$\\[5mm]\hline
        \\[-1.5em]
        $\E{\Sigma}{\Gamma}\left(\begin{array}{c}\infer[(O{\cdot}V\!ar)]  {\Gamma' \VDASH_{\Sigma'} x : \sigma} {\infer{\VDASHS \Gamma'}{\D} & x \of \sigma \in \Gamma'}\end{array} \right) \Longrightarrow \begin{array}{c}\infer[]  {\Gamma'' \VDASH_{\Sigma''} x : \sigma'} {\infer{\VDASH_{\Sigma''} \Gamma''}{\D'} & x \of \sigma' \in \Gamma''}\end{array} $\hfil  where $\E{\Sigma}{\Gamma}\left(\begin{array}{c}\infer{\VDASH_{\Sigma'} \Gamma'}{\D}\end{array}\right)  \Longrightarrow\begin{array}{c}\infer{\VDASH_{\Sigma''} \Gamma''}{\D'}\end{array}$\\[5mm]\hline
        \\[-1.5em]
        $\E{\Sigma}{\Gamma}\left(\begin{array}{c}\infer[(O{\cdot}Abs)]  {\Gamma' \VDASH_{\Sigma'} {\Abs x \sigma M} : {\Prod x \sigma \tau}}  {\infer{\Gamma', x \of \sigma \VDASH_{\Sigma'} M : \tau}{\D}}\end{array} \right) \Longrightarrow\begin{array}{c}\infer[] {\Gamma'' \VDASH_{\Sigma''} {\Abs x \sigma' M'} : {\Prod x \sigma' \tau'}}  {\infer{\Gamma'', x \of \sigma' \VDASH_{\Sigma''} M' : \tau'}{\D'}}\end{array}
        $\hfil
        \\[5mm]
        \hfill where $\E{\Sigma}{\Gamma}\left(\begin{array}{c}\infer{\Gamma', x \of \sigma \VDASH_{\Sigma'} M :  \tau}{\D}\end{array}\right) \Longrightarrow\begin{array}{c}\infer{\Gamma'', x \of \sigma' \VDASH_{\Sigma''} M' : \tau'}{\D'}\end{array}$\\[3mm]\hline
      \end{tabular}
    }
  \end{small}
\end{figure}

\begin{figure}
  \caption{Encoding of the ``standard'' object rules - Pt.2}\label{fig:obj_enc2}
  \begin{small}
    {\renewcommand{\arraystretch}{1.4}
      \begin{tabular}{|l|}\hline
        \\[-1.5em]
        $\E{\Sigma}{\Gamma}\left(\begin{array}{c}\infer[(O{\cdot}App)] {\Gamma' \VDASH_{\Sigma'} M \at N : \tau[N/x]} {\infer{\Gamma' \VDASH_{\Sigma'} M : {\Prod x \sigma \tau}}{\D_1} & \infer{\Gamma' \VDASH_{\Sigma'} N : \sigma}{\D_2}}\end{array} \right) \Longrightarrow \begin{array}{c}\infer {\Gamma'' \VDASH_{\Sigma^{iv}} M' \at N' : \tau'[N'/x]}{\infer{\Gamma'' \VDASH_{\Sigma^{iv}} M' : {\Prod x \sigma' \tau'}}{(\D'_1)^+} & \infer{\Gamma'' \VDASH_{\Sigma^{iv}} N' : \sigma'}{(\D'_2)^+}}\end{array} $\hfil
        \\[3mm]
        \hfill where $\E{\Sigma}{\Gamma}\left(\begin{array}{c}\infer{\Gamma' \VDASH_{\Sigma'} M : {\Prod x \sigma \tau}}{\D_1}\end{array}\right)  \Longrightarrow\begin{array}{c}\infer{\Gamma'' \VDASH_{\Sigma''} M' : {\Prod x \sigma' \tau'}}{\D'_1}\end{array}$, and \\[5mm]
        \hfill $\E{\Sigma}{\Gamma}\left(\begin{array}{c}\infer{\Gamma' \VDASH_{\Sigma'} N : \sigma}{\D_2}\end{array}\right) \Longrightarrow\begin{array}{c}\infer{\Gamma'' \VDASH_{\Sigma'''}  N' : \sigma'}{\D'_2}\end{array}$,  and  $\Sigma^{iv}\eqdef \M(\Sigma'',\Sigma''')$\\[3mm]\hline
        \\[-1.5em]
        $\E{\Sigma}{\Gamma}\left(\begin{array}{c}\infer[(\ast)]{\Gamma' \VDASH_{\Sigma'} M : \tau} {\infer{\Gamma' \VDASH_{\Sigma'} M : \sigma}{\D_1} & \infer{\Gamma' \VDASH_{\Sigma'} \tau : \Type}{\D_2} & \infer{\sigma \eqBL \tau}{\D_3}}\end{array} \right) \Longrightarrow \begin{array}{c}\infer[]{\Gamma^{iv} \VDASH_{\Sigma^{iv}} M' : \tau'} {\infer{\Gamma'' \VDASH_{\Sigma^{iv}} M' : \sigma'}{(\D'_1)^+} &\infer{\Gamma'' \VDASH_{\Sigma^{iv}} \tau' : \Type}{(\D'_2)^+} & \infer{\sigma' \eqB \tau'}{\D'_3}}\end{array}$\hfil\\[3mm]
        $(\ast) \eqdef (O{\cdot}Conv)$  \hfill where $\E{\Sigma}{\Gamma}\left(\begin{array}{c}\infer{\Gamma' \VDASH_{\Sigma'} M :\sigma}{\D_1}\end{array}\right) \Longrightarrow \begin{array}{c}\infer{\Gamma'' \VDASH_{\Sigma''} M' : \sigma'}{\D'_1}\end{array}$, and \\[3mm]
        \hfill $\E{\Sigma}{\Gamma}\left(\begin{array}{c}\infer{\Gamma' \VDASH_{\Sigma'} \tau : \Type}{\D_2}\end{array}\right) \Longrightarrow \begin{array}{c}\infer{\Gamma'' \VDASH_{\Sigma'''} \tau' : \Type}{\D'_2}\end{array}$,  and\\[3mm]
        \hfill $\Sigma^{iv}\eqdef \M(\Sigma'',\Sigma''')$,  and  $\E{\Sigma^{iv}}{\Gamma^{iv}}\left(\begin{array}{c}\infer{\sigma \eqBL \tau}{s\D_3}\end{array}\right) \Longrightarrow\begin{array}{c}\infer{\sigma' \eqB \tau'}{\D'_3}\end{array}$
        \\[3mm]\hline
      \end{tabular}
    }
  \end{small}
\end{figure}

\begin{figure}
  \caption{Encoding of the object rules involving locks and unlocks -
    Pt. 1}\label{fig:obj_lock_enc1}
  \begin{small}
    {\renewcommand{\arraystretch}{1.4}
      \begin{tabular}{|l|}\hline
        \\[-1.5em]
        $\E{\Sigma}{\Gamma}\left(\begin{array}{c}\infer[(O{\cdot}Lock)]{\Gamma' \VDASH_{\Sigma'} {\Lock {\P} N \sigma M} : {\Lock {\P} N \sigma {\rho}}} {\infer{\Gamma' \VDASH_{\Sigma'} M : \rho}{\D_1} & \infer{\Gamma' \VDASH_{\Sigma'} N : \sigma}{\D_2}}\end{array} \right) \Longrightarrow \begin{array}{c}\infer[] {\Gamma'' \VDASH_{\Sigma^{v}} \lambda y:(\P\ \vec{x}\ I_{\sigma'}\ N').M' : \Prod y {(\P\ \vec{x}\ I_{\sigma'}\ N')} \rho'} {\infer{\Gamma'' \VDASH_{\Sigma^{v}} M' : \rho'}{(\D'_1)^+} & \infer{\Gamma'' \VDASH_{\Sigma^{v}} N' : \sigma'}{(\D'_2)^+}}\end{array} $\hfil\\[7mm]
        \hfill where $\E{\Sigma}{\Gamma}\left(\begin{array}{c}\infer{\Gamma' \VDASH_{\Sigma'} M : \rho}{\D_1}\end{array}\right) \Longrightarrow\begin{array}{c}\infer{\Gamma'' \VDASH_{\Sigma''}M' : \rho'}{\D'_1}\end{array}$, and \\[3mm]
        \hfill $\E{\Sigma}{\Gamma}\left(\begin{array}{c}\infer{\Gamma' \VDASH_{\Sigma'} N : \sigma}{\D_2}\end{array}\right) \Longrightarrow\begin{array}{c}\infer{\Gamma'' \VDASH_{\Sigma'''} N' : \sigma'}{\D'_2}\end{array}$, and \\[3mm]
        \hfill $\Sigma^{iv}\eqdef \M(\Sigma'',\Sigma''')$, and $x_1,\ldots,x_n\eqdef \vec{x}\eqdef \Dom(\Gamma')$, and \\[3mm]
        \hfill $\Sigma^v=\left\{\begin{array}{lr}\Sigma^{iv} &\mbox{ if }\P\in \Dom(\Sigma^{iv})\\
        (\Sigma^{iv},\P \of \Pi x_1\of\sigma_1 \ldots x_n\of\sigma_n.(\sigma'\rightarrow \sigma')\rightarrow \sigma' \rightarrow\Type)& \mbox{ otherwise}
      \end{array}\right. $\\[5mm]\hline
        \\[-1.5em]
        $\E{\Sigma}{\Gamma}\left(\begin{array}{c}\infer[(\ast)] {\Gamma' \VDASH_{\Sigma'} {\Unlock {\P} N \sigma M} : \rho} {\infer{\Gamma' \VDASH_{\Sigma'} M : {\Lock {\P} N \sigma \rho}}{\D}  &  \P(\Gamma' \VDASH_{\Sigma'} N : \sigma)}\end{array} \right)\Longrightarrow \begin{array}{c}\infer[] {\Gamma''\VDASH_{\Sigma'''} (M'\at(c_{\P}\ \vec{x}\ I_{\sigma'}\ N')):\rho'} {\infer{\Gamma'' \VDASH_{\Sigma'''} M' : {\Prod y {(\P\ \vec{x}\ I_{\sigma'}\ N')}\ \rho'}}{(\D')^+}}\end{array}$\hfil\\[7mm]
       $(\ast)\eqdef (O{\cdot}Top{\cdot}Unlock)$  \hfill where $\E{\Sigma}{\Gamma}\left(\begin{array}{c}\infer{\Gamma' \VDASH_{\Sigma'} M : {\Lock {\P} N \sigma \rho}}{\D}\end{array}\right) \Longrightarrow\begin{array}{c}\infer{\Gamma'' \VDASH_{\Sigma''} M' : {\Prod y {(\P\ \vec{x}\ I_{\sigma'}\ N')}\ \rho'}}{\D'}\end{array}$, and \\[3mm]
        \hfill $x_1,\ldots,x_n \eqdef\vec{x}\eqdef \Dom(\Gamma')$, and \\[3mm]
        \hfill $\Sigma''' \eqdef  \left\{\begin{array}{lr} \Sigma'' & \mbox{ if }c_{P}\in \Dom(\Sigma'')\\ \Sigma'',c_{P} \of \Pi x_1\of \sigma_1\ldots x_n\of \sigma_n.\Pi x\of (\sigma'\rightarrow \sigma'),y\of\sigma'.(\P\ \vec{x}\ x \ y)) & \mbox{ otherwise} \end{array}\right.$\\[5mm] \hline
  \end{tabular}
}
\end{small}
\end{figure}

\begin{figure}
  \caption{Encoding of the $(F{\cdot}Guarded{\cdot}Unlock)$
    rule}\label{fig:fam_unlock_enc}
  \begin{small}
    {\renewcommand{\arraystretch}{1.4}
      \begin{tabular}{|l|}\hline
        \\[-1.5em]
        $\E{\Sigma}{\Gamma}\left(\begin{array}{c}\infer[(F{\cdot}Guarded{\cdot}Unlock)]  {\Gamma'  \VDASH_{\Sigma'} \Lock {\P} S \sigma { \rho[\Unlock {\P} {S'} {\sigma'} N /x]} : \Type} {\infer{\Gamma', x\of\tau \VDASH_{\Sigma'} \Lock {\P} S \sigma {\rho} : \Type}{\D_1} & \infer{\Gamma' \VDASH_{\Sigma'} N :  \Lock {\P} {S'} {\sigma'} \tau}{\D_2} & \infer{\sigma\eqBL\sigma'}{\D_3} & \infer{S\eqBL S'}{\D_4}}\end{array}  \right)$\\
        \hfil
        $\Longrightarrow$ \hfil\\
        \hfill $\left.\infer
        {\Gamma''\VDASH_{\Sigma^{iv}} \Pi y\of (\P\ \vec{x}\ {I_{\sigma''}}\ {S''}).\rho'[ N'  y/x] : \Type}  {{\infer[(T)]{\Gamma'', y\of (\P\ \vec{x}\ {I_{\sigma''}}\
        {S''}) \VDASH_{\Sigma^{iv}} \rho'[ N' y /x] : \Type}  {\infer{\phantom{\_\hspace{20em}\_}}{{\D''_1} \qquad {\D}}}}} \right. $\hfil\\[3mm]
        \hfill where $\D \eqdef \left.\begin{array}{l}{\infer{\Gamma'', y\of (\P\ \vec{x}\ {I_{\sigma''}}\ {S''})  \VDASH_{\Sigma^{iv}} N' y:\tau'}  {\infer[(w+\alpha)]{\infer{\Gamma'', y\of (\P\ \vec{x}\ {I_{\sigma''}}\ {S''})\VDASH_{\Sigma^{iv}} N' : \Pi y\of (\P\ \vec{x}\ {I_{\sigma''}}\    {S''}).\tau'}{}}  {\infer{\Gamma'' \VDASH_{\Sigma^{iv}} N' :\Pi y\of (\P\ \vec{x}\ {I_{\sigma''}}\    {S''}).\tau'}  {\infer{\Gamma'' \VDASH_{\Sigma^{iv}} N' : \Pi y\of (\P\ \vec{x}\ {I_{\sigma'''}}\   {S'''}).\tau'}{(\D'_2)^+} & \infer{\Pi y\of (\P\ \vec{x}\ {I_{\sigma'''}}\  {S'''}).\tau'\eqBL\Pi y\of (\P\ \vec{x}\   {I_{\sigma''}}\ {S''}).\tau'}  {\stackrel{\displaystyle\infer{\sigma''\eqB\sigma'''}{\D'_3} \ \ \infer{S''\eqB S'''}{\D'_4}}{\vdots}}}} \rew{3}  \deduce{\D_5}{\up{8}\mathrm{, and}} }}\end{array}\right.$ \\[3mm]
        \hfill $\D_5:\Gamma'', y\of (\P\ \vec{x}\ {I_{\sigma''}}\  {S''}) \VDASH_{\Sigma^{iv}}y:(\P\ \vec{x}\ {I_{\sigma''}}\  {S''})$,  and $\Delta\eqdef x\of\tau', y\of(\P\ \vec{x}\ {I_{\sigma''}}\ {S''})$, and \\[3mm]
        \hfill $(w+\alpha)$ stands for an application of weakening and $\alpha$-conversion in \LF, and \\[3mm]
        \hfill $(T)$ stands for an application of the transitivity theorem in \LF, and \\[3mm]
        \hfill vertical dots $\left(\vdots\right)$ stand for applications of context-closure and definitional equality rules, and \\[3mm]
        \hfill $\E{\Sigma}{\Gamma}\left(\begin{array}{c}\infer{\Gamma', x\of\tau \VDASH_{\Sigma'} \Lock {\P} S \sigma {\rho} : \Type}{\D_1}\end{array}\right)
        \Longrightarrow \begin{array}{c}\infer{\Gamma'', x\of\tau' \VDASH_{\Sigma''} \Pi y\of (\P\ \vec{x}\ {I_{\sigma''}}\ {S''}).\rho' : \Type}{\D'_1}\end{array}$, and \\[3mm]
        \hfill whence (for the Generation Lemma on Pure Type Systems~\cite{Bar-92}) there exists a derivation $\D''_1$\\[3mm]
        \hfill $\D''_1:\Gamma'', x\of\tau', y\of{(\P\ \vec{x}\ {I_{\sigma''}}\ {S''})}\VDASH_{\Sigma''} \rho' : \Type$, and \\[3mm]
        \hfill $\E{\Sigma''}{\Gamma''}\left(\begin{array}{c}\infer{\Gamma' \VDASH_{\Sigma'} N :  \Lock {\P} {S'} {\sigma'} \tau}{\D_2}\end{array}\right)
        \Longrightarrow\begin{array}{c}\infer{\Gamma''  \VDASH_{\Sigma'''} N' : \Pi y\of (\P\ \vec{x}\ {I_{\sigma'''}}\   {S'''}).\rho'}{\D'_2}\end{array}$, and \\[3mm]
        \hfill $\E{}{}\left(\begin{array}{c}\infer{\sigma\eqBL\sigma'}{\D_3}\end{array}\right) \Longrightarrow\begin{array}{c}\infer{\sigma''\eqB\sigma'''}{\D'_3}\end{array}$, and \\[3mm]
        \hfill $\E{}{}\left(\begin{array}{c}\infer{S\eqBL  S'}{\D_4}\end{array}\right) \Longrightarrow\begin{array}{c}\infer{S''\eqB S'''}{\D'_4}\end{array}$, and\\[3mm]
        \hfill $\Sigma^{iv}\eqdef \M(\Sigma'',\Sigma''')$, and $x_1,\ldots,x_n\eqdef \vec{x}\eqdef \Dom(\Gamma')$
        \\\hline
      \end{tabular}
    }
  \end{small}
\end{figure}

\begin{figure}
  \caption{Encoding of the object rules involving locks and unlocks -
    Pt. 2}\label{fig:obj_lock_enc2}
  \begin{small}
    {\renewcommand{\arraystretch}{1.4}
      \begin{tabular}{|l|}\hline
        \\[-1.5em]
        $\E{\Sigma}{\Gamma}\left(\begin{array}{c}\infer[(O{\cdot}Guarded{\cdot}Unlock)] {\Gamma'  \VDASH_{\Sigma'} {\Lock {\P} S \sigma {M[\Unlock {\P} {S'} {\sigma'} N /x]}} : \Lock {\P} S \sigma { \rho[\Unlock {\P} {S'} {\sigma'} N /x]}}  {\infer{\Gamma', x\of\tau \VDASH_{\Sigma'} \Lock {\P} S \sigma {M} : \Lock {\P} S \sigma {\rho}}{\D_1} \ \  \infer{\Gamma' \VDASH_{\Sigma'} N :  \Lock {\P} {S'} {\sigma'} \tau}{\D_2} \ \  \infer{\sigma\eqBL\sigma'}{\D_3}  \ \  \infer{S\eqBL S'}{\D_4}}\end{array} \right)$\\
        \hfil $\Longrightarrow$ \hfil\\
        $
        \hfill \left.\infer
        {\Gamma''\VDASH_{\Sigma^{iv}} \lambda y\of{(\P\ \vec{x}\ {I_{\sigma''}}\ {S''})}.M'[N'  y/x]:\Pi y\of (\P\ \vec{x}\ {I_{\sigma''}}\ {S''}).\rho'[N'  y/x]} {{\infer[(T)]{\Gamma'', y\of (\P\ \vec{x}\ {I_{\sigma''}}\ {S''}) \VDASH_{\Sigma^{iv}} M'[ N' y /x]:\rho'[ N' y/x]}{\infer{\phantom{\_\hspace{24em}\_}}{\infer{\Gamma'', \Delta \VDASH_{\Sigma^{iv}} M':\rho'}{\D} \qquad {\D'}}}}}
        \right.
        $\hfil\\[3mm]
        \hfill where
        $\D\eqdef \left.\begin{array}{l}{\infer[(SR)]{\Gamma'', \Delta \VDASH_{\Sigma^{iv}} M' : \rho'}  {\infer{\infer{}{\Gamma'', \Delta \VDASH_{\Sigma^{iv}} (\lambda y\of{(\P\ \vec{x}\ {I_{\sigma''}}\ {S''})}.M')y : \rho'} } {\infer[(w+\alpha)]{\infer{\Gamma'', \Delta \VDASH_{\Sigma^{iv}} \lambda y\of{(\P\ \vec{x}\ {I_{\sigma''}}\ {S''})}.M' : \Pi y\of (\P\ \vec{x}\ {I_{\sigma''}}\ {S''}).\rho'} {} } {(\D'_1)^+} & \deduce{\D_5}{\up{15}{\mathrm{, and}}} }}}  \end{array}\right.$ \\[3mm]
        \hfill $\D'\eqdef \left.\begin{array}{l}{\infer{\Gamma'', y\of (\P\ \vec{x}\ {I_{\sigma''}}\ {S''}) \VDASH_{\Sigma^{iv}} N' y : \tau'}{\infer[(w+\alpha)]{\infer{\Gamma'', y\of (\P\ \vec{x}\ {I_{\sigma''}}\ {S''})\VDASH_{\Sigma^{iv}} N' : \Pi y\of (\P\ \vec{x}\ {I_{\sigma''}}\ {S''}).\tau'}{}}{\infer{\Gamma'' \VDASH_{\Sigma^{iv}} N' : \Pi y\of (\P\ \vec{x}\ {I_{\sigma''}}\ {S''}).\tau'}{\infer{\Gamma'' \VDASH_{\Sigma^{iv}} N' : \Pi y\of(\P\ \vec{x}\ {I_{\sigma'''}}\ {S'''}).\tau'}{(\D'_2)^+} & \infer{\Pi y\of (\P\ \vec{x}\ {I_{\sigma'''}}\ {S'''}).\tau'\eqBL\Pi y\of(\P\ \vec{x}\ {I_{\sigma''}}\ {S''}).\tau'}{\stackrel{\displaystyle\infer{\sigma''\eqB\sigma'''}{\D'_3} \quad \infer{S''\eqB S'''}{\D'_4}}{\vdots}}}} \rew{3} \deduce{\D_6}{\deduce{\up{9}}{\mathrm{, and}}}}}\end{array}\right.$\\[3mm]
        \hfill $\D_5:\Gamma'', x\of\tau',y\of (\P\ \vec{x}\ {I_{\sigma''}}\ {S''}) \VDASH_{\Sigma^{iv}}y:(\P\ \vec{x}\ {I_{\sigma''}}\ {S''})$, and $\Delta\eqdef  x\of\tau', y\of(\P\ \vec{x}\ {I_{\sigma''}}\ {S''})$, and\\[3mm]
        \hfill $\D_6:\Gamma'', y\of (\P\ \vec{x}\ {I_{\sigma''}}\ {S''}) \VDASH_{\Sigma^{iv}}y:(\P\ \vec{x}\ {I_{\sigma''}}\ {S''})$, and\\[3mm]
        \hfill $(w+\alpha)$ stands for an application of weakening and $\alpha$-conversion in \LF,\\[3mm]
        \hfill $(SR)$ stands for an application of subject reduction in \LF, and \\[3mm]
        \hfill $(T)$ stands for an application of the transitivity theorem in \LF, and \\[3mm]
        \hfill vertical dots $\left(\vdots\right)$ stand for applications of context-closure and definitional equality rules, and \\[3mm]
        \hfill $\E{\Sigma}{\Gamma}\left(\begin{array}{c}\infer{\Gamma', x\of\tau \VDASH_{\Sigma'} \Lock {\P} S \sigma {M} : \Lock {\P} S \sigma {\rho}}{\D_1}\end{array}\right) \Longrightarrow \begin{array}{c}\infer{\Gamma'', x\of\tau' \VDASH_{\Sigma''} \lambda y\of (\P\ \vec{x}\ {I_{\sigma''}}\ {S''}) .M' : \Pi y\of (\P\ \vec{x}\ {I_{\sigma''}}\ {S''}).\rho'}{\D'_1 \fwd{60} \mathrm{,~and} }\end{array}$\\[3mm]
        \hfill $\E{\Sigma''}{\Gamma''}\left(\begin{array}{c}\infer{\Gamma' \VDASH_{\Sigma'} N  :  \Lock {\P} {S'} {\sigma'} \tau}{\D_2}\end{array}\right) \Longrightarrow\begin{array}{c}\infer{\Gamma''  \VDASH_{\Sigma'''} N' : \Pi y\of (\P\ \vec{x}\ {I_{\sigma'''}}\  {S'''}).\rho'}{\D'_2}\end{array}$, and \\[3mm]
        \hfill $\E{}{}\left(\begin{array}{c}\infer{\sigma\eqBL\sigma'}{\D_3}\end{array}\right) \Longrightarrow\begin{array}{c}\infer{\sigma''\eqB\sigma'''}{\D'_3}\end{array}$, and $\E{}{}\left(\begin{array}{c}\infer{S\eqBL S'}{\D_4}\end{array}\right) \Longrightarrow\begin{array}{c}\infer{S''\eqB S'''}{\D'_4}\end{array}$, and\\[3mm]
        \hfill $\Sigma^{iv}\eqdef \M(\Sigma'',\Sigma''')$, and $x_1,\ldots,x_n\eqdef \vec{x} \eqdef \Dom(\Gamma')$
        \\\hline
      \end{tabular}
    }
  \end{small}
\end{figure}
\begin{figure}\caption{Induced encoding of terms (kinds, families and objects)}\label{fig:terms_enc}
  {\renewcommand{\arraystretch}{1.4}
    \begin{tabular}{|lcl|}\hline
      $\ET{\Sigma}{\Gamma}(\Type)$ & $\eqdef$ &$\Type$\\[1mm]\hline
      $\ET{\Sigma}{\Gamma}({\Prod x \sigma K})$ & $\eqdef$ & $\Pi x\of \ET{\Sigma}{\Gamma}(\sigma). \ET{\Sigma}{\Gamma,x\of\ET{\Sigma}{\Gamma}(\sigma)}(K)$ \\[1mm]\hline
      $\ET{\Sigma}{\Gamma}(a)$ & $\eqdef$ & $a$ \hfill if $a\in\Dom(\Sigma)$\\[1mm]\hline
      $\ET{\Sigma}{\Gamma}({\Prod x \sigma \tau})$ & $\eqdef$ & $\Pi x\of \ET{\Sigma}{\Gamma}(\sigma).\ET{\Sigma}{\Gamma,x\of\ET{\Sigma}{\Gamma}(\sigma)}(\tau)$\\[1mm]\hline
      $\ET{\Sigma}{\Gamma}(\sigma N)$ & $\eqdef$ & $\ET{\Sigma}{\Gamma}(\sigma)\at \ET{\Sigma}{\Gamma}(N)$ \\[1mm]\hline
      $\ET{\Sigma}{\Gamma}({\Lock {\P} N \sigma \tau})$ & $\eqdef$ & $\Pi x\of (\P^{\Sigma}_{\Gamma}\ {\vec{x}}\ I_{\ET{\Sigma}{\Gamma}(\sigma)}\ {\ET{\Sigma}{\Gamma}(N)}).\ET{\Sigma}{\Gamma,x\of (\P^{\Sigma}_{\Gamma}\ {\vec{x}}\ I_{\ET{\Sigma}{\Gamma}(\sigma)}\ {\ET{\Sigma}{\Gamma}(N)})}(\tau)$\\ & & \hfill if $\P^\Sigma_\Gamma\in\Dom(\Sigma)$, and $\vec{x} \in \Dom(\Gamma)$\\[1mm]\hline
      $\ET{\Sigma}{\Gamma}(c)$ & $\eqdef$ & $c$  \hfill if $c\in\Dom(\Sigma)$ \\[1mm]\hline
      $\ET{\Sigma}{\Gamma}(x)$ & $\eqdef$ & $x$ \hfill if $x\in\Dom(\Gamma)$\\[1mm]\hline
      $\ET{\Sigma}{\Gamma}(\lambda x\of\tau.M)$ & $\eqdef$ & $\lambda x\of\ET{\Sigma}{\Gamma}(\tau).\ET{\Sigma}{\Gamma,x\of\ET{\Sigma}{\Gamma}(\tau)}(M)$ \\[1mm]\hline
      $\ET{\Sigma}{\Gamma}(MN)$ & $\eqdef$ & $\ET{\Sigma}{\Gamma}(M)\at \ET{\Sigma}{\Gamma}(N)$ \\[1mm]\hline
      $\ET{\Sigma}{\Gamma}({\Lock {\P} N \sigma M})$ & $\eqdef$ & $\lambda x\of(\P^{\Sigma}_{\Gamma}\ {\vec{x}}\ I_{\ET{\Sigma}{\Gamma}(\sigma)}\ \ET{\Sigma}{\Gamma}({N})).\ET{\Sigma}{\Gamma,x\of (\P^{\Sigma}_{\Gamma}\ {\vec{x}}\ I_{\ET{\Sigma}{\Gamma}(\sigma)}\ {\ET{\Sigma}{\Gamma}(N)})}(M)$\\  & &  \hfill if $\P^\Sigma_\Gamma\in\Dom(\Sigma)$, and $\vec{x} \in \Dom(\Gamma)$)\\[1mm]\hline
      \\[-1.5em]
      $\ET{\Sigma}{\Gamma}({\Unlock {\P} N \sigma M})$ & $\eqdef$ & $\left\{\begin{array}{l}  \ET{\Sigma}{\Gamma}({M}) \at x  \quad \hfill \mbox{if }\P^{\Sigma}_{\Gamma}\in \Dom(\Sigma)$, and  $x\of(\P^{\Sigma}_{\Gamma}\ {\vec{x}}\ I_{\ET{\Sigma}{\Gamma}(\sigma)}\ {\ET{\Sigma}{\Gamma}(N)})\in\Gamma\\ \ET{\Sigma}{\Gamma}(M) \at  (c_{\P^\Sigma_\Gamma}\ {\vec{x}}\ I_{\ET{\Sigma}{\Gamma}(\sigma)}\ {\ET{\Sigma}{\Gamma}(N)}) \hfill   \mbox{if }c_{\P^\Sigma_{\Gamma}},\P^{\Sigma}_{\Gamma}\in \Dom(\Sigma)\end{array}\right.$\\[5mm]\hline
    \end{tabular}
  }
\end{figure}
Starting from signatures and contexts, we have the encoding clauses of
Figures~\ref{fig:sig_enc} and~\ref{fig:ctx_enc}.  In particular,
notice how rules having two premises, \ie\ rules ($S{\cdot}Kind$),
($S{\cdot}Type)$, and ($C{\cdot}Type$), are dealt with. In encoding
locks in \LF\ we extend the signature, therefore we have to reflect on
the encoding of the derivation of the first premise the effects of
encoding the second premise and viceversa.

In Figures~\ref{fig:kind_enc}, \ref{fig:family_enc1},
\ref{fig:family_enc2}, \ref{fig:obj_enc1}, \ref{fig:obj_enc2},
\ref{fig:obj_lock_enc1}, \ref{fig:fam_unlock_enc}
and~\ref{fig:obj_lock_enc2} appear the clauses defining the encoding
of derivations concerning terms (\ie, kinds, families, and
objects). The \emph{key clause} of our encoding is ($F{\cdot}Lock$),
mapping a lock type in \LLFP\ to a $\Pi$-type in \LF:
$$
{\Lock {\P} N \sigma \rho} \leadsto \Pi y\of (\P^{\Sigma}_{\Gamma}\
  \vec{x}\ I_{\overline{\sigma}}\ \overline{N}).\overline{\rho}
$$
Correspondingly at the level of objects we have the following key
steps (where $\Sigma'$ and $\Gamma'$ are, respectively, the signature
and the typing context in \LF\ coming from the corresponding encoding
of the derivation):
$$
\begin{array}{lcl}
  {\Lock {\P} N \sigma M} & \leadsto  & \left\{\begin{array}{ll}\lambda  x\of \pred{\Sigma}{\Gamma}{\vec{x}}{\sigma}{N}.\overline{M}  & \qquad\rew{1} \mbox{if }\P^\Sigma_\Gamma\in\Dom(\Sigma')\end{array}\right.\\[2mm]  {\Unlock {\P} N \sigma M} & \leadsto & \left\{\begin{array}{ll} (\overline{M}\at x) & \qquad \mbox{if }\P^{\Sigma}_{\Gamma}\in \Dom(\Sigma') \mbox{ and }x \of \pred{\Sigma}{\Gamma}{\vec{x}}{\sigma}{N}\in\Gamma'\\[2mm] (\overline{M}\at \termP{\Sigma}{\Gamma}{\vec{x}}{\sigma}{N}) &\qquad \mbox{if }c_{\P^\Sigma_{\Gamma}}, \P^{\Sigma}_{\Gamma}\in \Dom(\Sigma')\end{array}\right.
\end{array}
$$
Of course, this amounts to a form of \emph{proof irrelevance}, as it
should be, since in \LLFP\ the verification of
$\P({\Gamma\VDASHS N : \sigma})$ is carried out by an external system
(\ie, the \emph{oracle}) which only returns a yes/no answer. Hence the
external proof argument is not important and it can be represented by
a constant or variable. In the remaining cases, the encoding function
behaves in a straightforward way propagating itself to inner
subderivations, \eg\ in the rules of application, abstraction etc.
\enlargethispage{\baselineskip}

As far as the typing rules, we point out the possible extension of the
signature in the encoding of rules $(F{\cdot}Lock)$
(Figure~\ref{fig:family_enc2}) and $(O{\cdot}Lock)$
(Figure~\ref{fig:obj_lock_enc1}) when a, possibly new, lock constant
is introduced and in rule $(O{\cdot}Top{\cdot}Unlock)$
(Figure~\ref{fig:obj_lock_enc1}) when a term is unlocked by the
external oracle call.
%
%
Finally we can establish the following crucial theorem concerning the
encoding functions $\E{}{}$ and $\ET{\Sigma}{\Gamma}$:

\begin{thm}[Properties of $\E{\Sigma}{\Gamma}$ and
  $\ET{\Sigma}{\Gamma}$]\label{th:enc_prop} The encoding function
  $\E{\Sigma}{\Gamma}$ defined in Figures~\ref{fig:sig_enc},
  \ref{fig:ctx_enc}, \ref{fig:kind_enc}, \ref{fig:family_enc1},
  \ref{fig:family_enc2}, \ref{fig:obj_enc1}, \ref{fig:obj_enc2},
  \ref{fig:obj_lock_enc1}, \ref{fig:fam_unlock_enc},
  and~\ref{fig:obj_lock_enc2} satisfies the following properties:
  \begin{enumerate}
  \item $\E{\Sigma}{\Gamma}$ encodes derivations coherently, \ie,
    derivations of signatures/contexts well-formed\-ness in \LLFP\ are
    mapped to derivations of signatures/contexts well-formedness in
    \LF, typing derivations in \LLFP\ are mapped to typing derivations
    in \LF, etc.
  \item $\E{}{}$ is a total compositional function.
  \item Given a valid derivation $\D : \Gamma\VDASHS A:B$ in \LLFP,
    then the derivation, denoted by $\E{\emptyset}{\emptyset}(\D) :
    \ 
    \overline{\Gamma}\VDASH_{\overline{\Sigma}}\overline{A}:\overline{B}$,
    is a valid derivation in~\LF.
  \item Given a valid derivation $\D
    : A\evalBL B$ (resp. $\D : A\eqBL
    B$) in \LLFP, then the derivation, denoted by
    $\E{\emptyset}{\emptyset}(\D) :
    \ 
    \overline{A}\evalB\overline{B}$
    (resp. $\E{\emptyset}{\emptyset}(\D) :
    \ 
    \overline{A}\eqB\overline{B}$), is a valid derivation in~\LF.
  \item The compositional function $\E{}{}$
    induces the function $\ET{\Sigma}{\Gamma}$
    between terms (kinds, families and objects) of \LLFP\ and of \LF\
    defined in Figure~\ref{fig:terms_enc}. The signature $\Sigma$
    and the typing context $\Gamma$
    passed as parameters to the induced map are precisely the final
    ones generated by the translation process of the derivations.
  \item If $\D$
    is a valid derivation in \LLFP\ of the typing judgment
    $\Gamma\VDASHS
    A : B$, then $\E{}{}(\D: \Gamma\VDASHS A : B) = \D' :
    \overline{\Gamma} \VDASH_{\overline{\Sigma}}
    \ET{\overline{\Gamma}}{\overline{\Sigma}}(A)
    :\ET{\overline{\Gamma}}{\overline{\Sigma}}(B)$.
  \item If the judgments $\Gamma\VDASHS
    A:B$ and $A\evalBL A'$ (resp. $A\eqBL
    A'$) hold in \LLFP, then we have
    $\ET{\overline{\Sigma}}{\overline{\Gamma}}(A)\evalB\ET{\overline{\Sigma}}{\overline{\Gamma}}(A')$
    (resp.
    $\ET{\overline{\Sigma}}{\overline{\Gamma}}(A)\eqB\ET{\overline{\Sigma}}{\overline{\Gamma}}(A')$)
    in \LF.
  \end{enumerate}

  \Proof: By a tedious but ultimately straightforward induction on
  derivations in \LLFP, taking into account the clauses of
  $\E{\Sigma}{\Gamma}$ and $\ET{\Sigma}{\Gamma}$.

\end{thm}
The judgements generated by the encoding \E{}{} make a very
special use of variables of type
$\pred{\Sigma}{\Gamma}{\vec{x}}{\sigma}{N}$.
Namely such variables can $\lambda$-bind only objects and $\Pi$-bind
only types, \ie\ terms of kind {\tt Type}. In particular families are
never $\lambda$-abstracted and kinds are never $\Pi$-bound by such
variables. Furthermore such variables are introduced in the encoding
only in a controlled way by the lock-unlock rules. This is crucial to
preserve the reversibility of the encoding necessary for transferring
the properties from \LF\ to \LLFP.
Notice that the terms produced by the encoding \ET{\Sigma}{\Gamma} satisfy similar
properties.  This kind of closure property, which we call
\emph{goodness}, is defined as follows:


\begin{defi}[Good Judgements and good terms]
  We call \emph{good judgements} the \LF-judgements whose subterms of
  type $\pred{\Sigma}{\Gamma}{\vec{x}}{\sigma}{N}$ (for some $\sigma$
  and $N$ in \LLFP) are only terms of the shape
  $\termP{\Sigma}{\Gamma}{\vec{x}}{\sigma}{N}$ or variables, which
  moreover appear as proper subterms always in applied
  position. Moreover only object terms are $lambda$-bound by variables
  of type $\pred{\Sigma}{\Gamma}{\vec{x}}{\sigma}{N}$ (for some
  $\sigma$ and $N$ in \LLFP) and only type terms, \ie\ objects of kind
  {\tt Type}, are $\Pi$-bounded over variables of type
  $\pred{\Sigma}{\Gamma}{\vec{x}}{\sigma}{N}$ (for some $\sigma$ and
  $N$ in \LLFP).  Terms occurring in good judgements are called
  \emph{good terms}.
\end{defi}
The introduction of good judgements is motivated by the following:
\begin{thm}[Good judgements]\label{th:good_terms}
  The \LF-judgements in the codomain of \E{}{} are good judgements.

  \Proof: By induction on the definition of \E{}{}, one can easily
  check that since whenever variables, or terms, of the shape
  $\termP{\Sigma}{\Gamma}{\vec{x}}{\sigma}{N}$ are introduced these
  are always applied to the translation of the body of an unlock-term
  in \LLFP.
\end{thm}
Moreover, $\beta$-reductions carried out in \LF\ preserve the
``goodness'' property. Indeed, if the term in functional position is a
$\lambda$-term, it will either erase the variable/term (representing
an unlock/lock dissolution) or replace it into another term, again, in
argument position. This latter case arises in derivations involving
the $(O\cdot Guarded\cdot Unlock)$-rule and the
$(F\cdot Guarded\cdot Unlock)$-rule in \LLFP.  This remark leads
immediately to the following proposition:

\begin{thm}[Closure by reduction of good terms]
  Good terms are closed by $\beta$-reduction in \LF, \ie, if $M$ is
  good and $M\rightarrow_{\beta} M'$ in \LF, then also $M'$ is a good
  term.
\end{thm}

To be able to transfer back to \LLFP\ properties of \LF\ we need to
``invert'' the encoding function $\E{}{}$.  This can be done only
starting from a good judgment in \LF. To this end we establish the
following proposition.

\begin{thm}
  If $\Gamma\VDASHS M:\sigma$ is a good \LF\ judgement, then there is
  a derivation of that judgment in \LF, all whose judgements are good
  \LF\ judgements.

  \Proof By induction on derivations in \LF. The only critical cases
  being type equality rules, which are nonetheless straightforward.
\end{thm}
Given the \E{}{}-encoding of derivations and the induced
\ET{\Sigma}{\Gamma}-encoding defined in Figure~\ref{fig:terms_enc}, we
can define an inverse function \DT{\Sigma}{\Gamma} on terms (see
Figure~\ref{fig:terms_dec}), where $\Sigma$ and $\Gamma$ are,
respectively the signature and the typing context of the corresponding
typing judgment in \LF. The following fundamental \emph{invertibility}
theorem establishes the above claim more precisely. This theorem will
be the main tool for transferring the metatheoretic properties of \LF\
to \LLFP. Notice that it amounts to a form of \emph{adequacy} of the
encoding $\E{}{}$.

\begin{thm}[Invertibility]\label{th:enc_inv}
  There exists a total function $\Einv$ mapping derivations of good
  judgements in \LF\ into derivations of \LLFP.  More precisely if
  $\D$ is a valid derivation in \LF\ of the \emph{good judgment}
  $\Gamma\VDASHS A : B$, then
  $\Einv(\D: \Gamma\VDASHS A : B) = \D' : \Gamma' \VDASH_{\Sigma'}
  \DT{\Gamma}{\Sigma}(A) :\DT{\Gamma}{\Sigma}(B)$,
  where the function \DT{\Sigma}{\Gamma} between terms (kinds,
  families and objects)
  of \LF\ and terms of \LLFP\ is defined in
  Figure~\ref{fig:terms_dec}.  In particular the function
  \DT{\Sigma}{\Gamma} is left inverse to \ET{\Sigma}{\Gamma}, \ie, for
  each derivation term $M$ such that $\Gamma\VDASHS M:\tau$ in \LLFP,
  we have that
  $\DT{\overline{\Sigma}}{\overline{\Gamma}}(\ET{\overline{\Sigma}}{\overline{\Gamma}}(M))=M$
  holds.  \Proof (Sketch): The definition of $\D$ is syntax-driven by
  the structure of \LF\ terms which are also \emph{good terms} (see
  Theorem~\ref{th:good_terms}). The only critical cases are the
  following:
\begin{enumerate}
\item application: it can represent an \emph{ordinary} application in
  \LLFP\ if the argument is not of the shape $(c_{\P}\ldots)$. The
  case of a variable cannot arise at ``top level'' since such variable
  is explained away making use of a \emph{guarded unlock}. Instead, if
  the argument is of the shape $(c_{\P}\ldots)$, \ie, a constant of
  type $(\P\ldots)$, then we can invert the derivation by means of a
  $(O{\cdot}Top{\cdot}Unlock)$-rule.
\item abstraction introduction: it can lead to an \emph{ordinary}
  abstraction in \LLFP\ if the type of the abstracted variable is not
  of the shape $(\P\ldots)$. Otherwise we have to introduce either a
  \emph{lock} or a \emph{guarded unlock}, depending on the occurrences
  of the abstracted variable of type $(\P\ldots)$ in the abstraction's
  body. Indeed, if there are no occurrences of the free variable, \ie,
  we have a \emph{dummy} abstraction, we invert it by a simple lock
  introduction; on the other hand, if there are one or more
  occurrences of the variable, we are in the second subcase.  Indeed,
  let us suppose we have the following derivation in \LF:

  \begin{eqnarray}\label{formula:lambda_intro}
    \infer{\Gamma\VDASHS \lambda y\of(\P\ \vec{x}\ I_{\sigma}\ S).M :
    \Pi y\of(\P\ \vec{x}\ I_{\sigma}\ S).\rho}
    {\infer{\Gamma, y\of(\P\ \vec{x}\ I_{\sigma}\ S)\VDASHS M:\rho}{\D}}
  \end{eqnarray}
  \noindent Then, if $y$ occurs in $M$, it must occur in argument
  position since we are working with \emph{good terms}, \ie, as a
  subterm of $Ny:\tau$ for a suitable $N$ and $\tau$, \ie,
  $M\equiv M^*[Ny]$ (where $M^*[\cdot]$ denotes a \emph{context}, \ie,
  term with a ``hole'' which can be filled by another term). For the
  sake of simplicity we discuss first the case in which all
  occurrences of y are applied to the same term $N$.  From above we
  can define $M'\equiv M[x/N y]$ and $\rho'\equiv \rho[x/N y]$ for a
  suitable $x$ of type $\tau$ such that $x\not\in\Dom(\Gamma)$ and
  $x\not\equiv y$. Thus, we can infer the existence of a derivation
  $\D':\Gamma, y\of(\P\ \vec{x}\ I_{\sigma}\ S), x\of\tau\VDASHS
  M':\rho'$.
  Moreover, there must be a derivation
  $\D_2: \Gamma, y\of(\P\ \vec{x}\ I_{\sigma}\ S)\VDASHS Ny:\tau$ from
  $\D_3:\Gamma, y\of(\P\ \vec{x}\ I_{\sigma}\ S) \VDASHS N:\Pi
  y\of(\P\ \vec{x}\ I_{\sigma'}\ {S'}).\tau$,
  and
  $\D_4:\Gamma, y\of(\P\ \vec{x}\ I_{\sigma}\ S)\VDASHS y\of(\P\
  \vec{x}\ I_{\sigma}\ S)$,
  and $\D_5: S\eqB S'$, and $\D_6: \sigma\eqB \sigma'$.

  \noindent Therefore we can perform the following proof
  ``manipulation'':

$$
\begin{array}{c}
  \infer{\Gamma\VDASHS \lambda y\of(\P\ \vec{x}\ I_{\sigma}\ S).M :
  \Pi y\of(\P\ \vec{x}\ I_{\sigma}\ S).\rho} {\infer{\Gamma, y\of(\P\ \vec{x}\ I_{\sigma}\ S)\VDASHS M:\rho}{\D}}\\
  \leadsto\\
  \infer{\Gamma\VDASHS \lambda y\of(\P\ \vec{x}\ I_{\sigma}\ S).M'[ N y /x] :  \Pi y\of(\P\ \vec{x}\ I_{\sigma}\ S).\rho'[ N y /x]}{\infer[(transitivity)]{\Gamma, y\of(\P\ \vec{x}\ I_{\sigma}\ S)\VDASHS M'[ N y /x]:\rho'[ N y /x]} {\infer{\phantom{\hspace{22em}\_}}{\D' & \infer{\Gamma, y\of(\P\ \vec{x}\ I_{\sigma}\ S)\VDASHS Ny:\tau}{\infer{\Gamma, y\of(\P\ \vec{x}\ I_{\sigma}\ S)\VDASHS N:\Pi y\of(\P\ \vec{x}\ I_{\sigma}\ {S}).\tau}{\D_3 & \infer{\Pi y\of(\P\ \vec{x}\ I_{\sigma'}\ {S'}).\tau\eqB \Pi y\of(\P\ \vec{x}\ I_{\sigma}\ {S}).\tau}{\stackrel{\D_5 \quad\D_6}{\vdots}}} \quad \D_4}}}}
\end{array}
$$
\medskip

\noindent Obviously, $M'[ N y /x]\equiv M[x/ N y ][ N y /x]\equiv M$
and $\rho'[N y /x]\equiv \rho[x/ N y ]$ $[ N y /x]\equiv \rho$.
Whence, we can define the following ``decoding'' of the abstraction
introduction~(\ref{formula:lambda_intro}), where $\D'_3$ is
essentially the derivation
$$\D_3:\Gamma, y\of(\P\ \vec{x}\ I_{\sigma}\ S)\VDASHS N:\Pi y\of(\P\
\vec{x}\ I_{\sigma'}\ {S'}).\tau$$
with a final application of \emph{strengthening} to prune the variable
$y$ from the typing context:

$$
\begin{array}{c}
  \infer{\Gamma'\VDASH_{\Sigma'}{\Lock{\P}{S''}{\sigma''}{M''[\Unlock{\P}{S'''}{\sigma'''}{N'}/x]}}:{\Lock{\P}{S''}{\sigma''}{\rho''[\Unlock{\P}{S'''}{\sigma'''}{N'}/x]}}} {\Einv \left(\begin{array}{c}\infer{\Gamma, x\of\tau\VDASHS \lambda y\of(\P\ \vec{x}\ I_{\sigma}\ S).M':\Pi y\of(\P\ \vec{x}\ I_{\sigma}\ S).\rho'}{\infer{\Gamma, x\of\tau, y\of(\P\ \vec{x}\ I_{\sigma}\ S)\VDASHS M':\rho'}{\D'}}\end{array}\right) & \Einv\left(\D'_3\right) & \Einv\left(\D_5\right) & \Einv\left(\D_6\right)}\\[3mm]
  \Longrightarrow\\[3mm]
  \infer{\Gamma'\VDASH_{\Sigma'}{\Lock{\P}{S''}{\sigma''}{M''[\Unlock{\P}{S'''}{\sigma'''}{N'}/x]}}:{\Lock{\P}{S''}{\sigma''}{\rho''[\Unlock{\P}{S'''}{\sigma'''}{N'}/x]}}} {\infer{\Gamma', x\of\tau'\VDASH_{\Sigma'} \Lock{\P}{S''}{\sigma''}{M''}:\Lock{\P}{S''}{\sigma''}{\rho''}} {\Einv\left(\D''\right)}\quad\infer{\Gamma'\VDASH_{\Sigma'} N':\Lock{\P}{S'''}{\sigma'''}{\tau'}}{\Einv\left(\D''_3\right)} & \infer{S''\eqBL S'''}{\Einv(\D'_5)} & \infer{\sigma''\eqBL\sigma'''}{\Einv(\D'_6)}}
\end{array}
$$
\medskip

\noindent where $\D''$, $\D''_3$, $\D'_5$ and $\D'_6$ are,
respectively, the residuals of derivations $\D'$, $\D'_3$, $\D_5$ and
$\D_6$ once we have got rid of the last applied rule.
\noindent The signature $\Sigma'$ and the typing context $\Gamma'$ in
\LLFP\ are obtained from the corresponding entities $\Sigma$ and
$\Gamma$ in \LF\ by pruning all the constants $\P$ and $c_{\P}$ and
variables of type $(\P\ldots)$.

\noindent The case in which the bound variable $y$ occurs as argument
of several different $N_1$,\ldots,$N_k$, the above decoding has to be
carried out repeatedly for each $N_i$.

\end{enumerate}

\noindent The fact that the decoding via \Einv\ of the derivation in
\LF\ of the typing judgment
$\Gamma\VDASH_{\Sigma}\lambda y\of(\P\ \vec{x}\ I_{\sigma}\ S).M'[ N
y/x]:\Pi y\of(\P\ \vec{x}\ I_{\sigma}\ S).\rho'[ N y /x]$
corresponds to a derivation in \LLFP\ of type
$\Gamma'\VDASH_{\Sigma'}\DT{\Sigma}{\Gamma}(\lambda y\of(\P\ \vec{x}\
I_{\sigma}\ S).M'[ N y /x]):\DT{\Sigma}{\Gamma}(\Pi y\of(\P\ \vec{x}\
I_{\sigma}\ S).\rho'[N y/x])$
can be easily verified, by looking at the defining clauses in
Figure~\ref{fig:terms_dec} and exploiting the fact that
$\DT{\Sigma}{\Gamma}(M[N/x])=\DT{\Sigma}{\Gamma}(M)[\DT{\Sigma}{\Gamma}(N)/x]$.

\noindent All the properties of the decoding function $\DT{}{}{}$ are
proved by induction taking into account the fact that all terms are
good.
\end{thm}

\begin{figure}\caption{Induced decoding of terms (kinds, families and objects)}\label{fig:terms_dec}
  {\renewcommand{\arraystretch}{1.4}
    \begin{tabular}{|lcl|}\hline
      $\DT{\Sigma}{\Gamma}(\Type)$ & $\eqdef$ & $\Type$\\[1mm]\hline
      $\DT{\Sigma}{\Gamma}({\Prod x \sigma K})$ & $\eqdef$ &  $\Pi x\of \DT{\Sigma}{\Gamma}(\sigma).\DT{\Sigma}{\Gamma,x\of\DT{\Sigma}{\Gamma}(\sigma)}(K)$\\[1mm]\hline
      $\DT{\Sigma}{\Gamma}(a)$ & $\eqdef$ & $a$ \hfill if $a\in\Dom(\Sigma)$, and $a\not \equiv\P^{\Sigma'}_{\Gamma'}$\\[1mm]\hline
      $\DT{\Sigma}{\Gamma}({\Prod x \sigma \tau})$ & $\eqdef$ &  $\left\{\begin{array}{lr} {\Lock \P {\DT{\Sigma}{\Gamma}(N)} {\DT{\Sigma}{\Gamma}(\sigma')} {\DT{\Sigma}{\Gamma,x:\sigma}(\tau)}}  & \hfill \fwd{33}\mbox{ if }\sigma\equiv(\P^{\Sigma'}_{\Gamma'}\ \vec{x}\ I_{\sigma'}\ N)\\[2mm] \Pi x\of\DT{\Sigma}{\Gamma}(\sigma).\DT{\Sigma}{\Gamma,x\of\DT{\Sigma}{\Gamma}(\sigma)}(\tau) & \hfill \mbox{ otherwise}
\end{array}\right.$\\[7mm]\hline
      $\DT{\Sigma}{\Gamma}(\sigma N)$ & $\eqdef$ & $\DT{\Sigma}{\Gamma}(\sigma)\at \DT{\Sigma}{\Gamma}(N)$ \\[1mm]\hline
      $\DT{\Sigma}{\Gamma}(c)$ & $\eqdef$ & $c$ \hfill if $c\in\Dom(\Sigma)$, and $c\neq c_{\P^{\Sigma'}_{\Gamma'}}$)\\[1mm]\hline
      $\DT{\Sigma}{\Gamma}(x)$ & $\eqdef$ & $x$ \hfill if $x\in\Dom(\Gamma)$, and the type of $x$ is not of the form $(\P^{\Sigma'}_{\Gamma'}\ \vec{x}\ I_{\sigma'}\ N)$\\[1mm]\hline
      $\DT{\Sigma}{\Gamma}(\lambda x\of\tau.M)$ & $\eqdef$ & $\left\{\begin{array}{l} {\Lock{\P}{\DT{\Sigma}{\Gamma}(N)}{\DT{\Sigma}{\Gamma}(\sigma')}{\DT{\Sigma}{\Gamma,x\of\tau}(M)}} \hfill  \fwd{34} \mbox{ if }\tau\equiv(\P^{\Sigma'}_{\Gamma'}\ \vec{x}\ I_{\sigma'}\ N)\\[2mm] \lambda x\of\DT{\Sigma}{\Gamma}(\tau).\DT{\Sigma}{\Gamma,x\of\DT{\Sigma}{\Gamma}(\tau)}(M) \hfill \mbox{ otherwise}\end{array}\right.$
      \\[7mm]\hline
      $\DT{\Sigma}{\Gamma}(MN)$ & $\eqdef$ & $\left\{\begin{array}{l} {\Unlock{\P}{\DT{\Sigma}{\Gamma}(N')}{\DT{\Sigma}{\Gamma}(\sigma)}{\DT{\Sigma}{\Gamma}(M)}} \hfill \mbox{ if } N\equiv(c_{\P^\Sigma_\Gamma}\ {\vec{x}}\ I_{\sigma}\ {N'})\\[2mm] {\Unlock{\P}{\DT{\Sigma}{\Gamma}(N')}{\DT{\Sigma}{\Gamma}(\sigma)}{\DT{\Sigma}{\Gamma}(M)}} \hfill  \fwd{18} \mbox{ if }N\equiv y\mbox{, and }y\of ({\P^\Sigma_\Gamma}\ {\vec{x}}\ I_{\sigma}\ {N'})\in\Gamma\\[2mm] \DT{\Sigma}{\Gamma}(M)\at\DT{\Sigma}{\Gamma}(N)\hfill \mbox{ otherwise}\end{array}\right.$ \\[10mm]\hline
    \end{tabular}
  }
\end{figure}


\section{Metatheory of \LLFP}\label{sec:metath}
We are now ready to prove all the most significant metatheoretic
properties of \LLFP. As we remarked earlier, most of them, and most
notably \emph{Strong Normalization} and \emph{Subject Reduction}, can
be inherited uniformly from those of \LF\ using
Theorems~\ref{th:enc_prop} and~\ref{th:enc_inv} and the other results
of the previous section.

\subsection{Confluence}
As it is often the case for systems without $\eta$-like conversions,
confluence can be proved directly on \emph{raw terms} as
in~\cite{HHP-92,honsell:hal-00906391}. Namely using \emph{Newman's
  Lemma} (\cite{Bar-book}, Chapter 3), and showing that the reduction
on ``raw terms'' is \emph{locally confluent}. But we can also make use
of the decoding function $\DT{}{}{}$ on closed good terms. Hence, we
have:
\begin{thm}[Confluence of \LLFP]
  \label{thm:confluence}
  $\beta\mathcal{L}$-reduction is confluent, \ie:
    \begin{enumerate}
    \item If $K \multievalBL K'$ and $K \multievalBL K''$, then there
      exists a $K'''$ such that $K' \multievalBL K'''$ as well as
      $K'' \multievalBL K'''$.
    \item If $\sigma \multievalBL \sigma'$ and
      $\sigma \multievalBL \sigma''$, then there exists a $\sigma'''$
      such that $\sigma' \multievalBL \sigma'''$ as well as
      $\sigma'' \multievalBL \sigma'''$.
    \item If $M \multievalBL M'$ and $M \multievalBL M''$, then there
      exists an $M'''$ such that $M' \multievalBL M'''$ as well as
      $M'' \multievalBL M'''$.
    \end{enumerate}
\end{thm}

\subsection{Strong Normalisation}
\noindent Strong normalisation can be proved following the same
pattern used in~\cite{honsell:hal-00906391}, relying on the strong
normalization of \LF, as proven in~\cite{HHP-92}. However, we do not
use a suitable extension of the ``forgetful'' function
$^\Erase:\LLFP \rightarrow \LF$ (introduced
in~\cite{honsell:hal-00906391}), which maps $\LLFP$ terms into $\LF$
terms essentially deleting the $\L$ and $\U$ symbols, but rather we
use the very encoding functions introduced in
Section~\ref{sec:LLFP_in_LF}.  Consider a well typed term $M$ of
\LLFP\ such that $\Gamma\VDASHS M:\sigma$. Without loss of generality
we can assume that $M$ is closed. It is immediate to check that any
sequence of $\evalBL$-reductions starting from $M$ can be reflected in
an $\evalB$ reduction starting from $\ET{\Sigma}{\Gamma}(T)$ of the
same length. This is a serendipitous consequence of the choice of
encoding ``locks as abstractions'' and ``unlocks as
applications''. Therefore, an infinite $\evalBL$-reduction in \LLFP\
would produce an infinite $\evalB$-reduction in \LF, which is
impossible, because \LF\ is strongly normalizing.

\begin{thm}[Strong normalization of \LLFP] \hfill
\label{thm:LFP_strnorm}
  	\begin{enumerate}
  	\item If $\Gamma\VDASHS K$, then $K$ is $\evalBL$-strongly
          normalizing.
  	\item if $\Gamma\VDASHS \sigma: K$, then $\sigma$ is
          $\evalBL$-strongly normalizing.
  	\item \label{osn} if $\Gamma\VDASHS M:\sigma$, then $M$ is
          $\evalBL$-strongly normalizing.
  	\end{enumerate}
\end{thm}

\subsection{Subject Reduction}
Using Theorems~\ref{th:enc_prop} and~\ref{th:enc_inv}, and in
particular the property concerning the interplay between $\E{}{}$ and
$\ET{\Sigma}{\Gamma}$ and the one between $\Einv$ and
$\DT{\Sigma}{\Gamma}$, it is easy to argue that the inversion function
$\Einv$ commutes with reduction, \ie\ inverting into \LLFP\ the reduct
in \LF\ produces the reduct in \LLFP.

Whence, we can deduce the fundamental theorem of subject reduction:

\begin{thm}[Subject reduction of \LLFP] If predicates are
  well-behaved, then:
\label{thm:subred}
\begin{enumerate}
\item If $\Gamma \VDASHS K$, and $K \evalBL K'$, then
  $\Gamma \VDASHS K'$.
\item If $\Gamma \VDASHS \sigma : K$, and $\sigma \evalBL \sigma'$,
  then $\Gamma \VDASHS \sigma' : K$.
\item If $\Gamma \VDASHS M : \sigma$, and $M \evalBL M'$, then $\Gamma
  \VDASHS M' : \sigma$.
\end{enumerate}

\Proof (sketch): Let us assume to have a derivation
$\D_1:\Gamma \VDASHS M : \sigma$, and a reduction for $M \evalBL M'$
in \LLFP. Then we encode $\D_1$ with $\E{}{}$, yielding
$\D'_1:\Gamma' \VDASH_{\Sigma'} M'' : \sigma'$, and the terms with
$\ET{\Sigma'}{\Gamma'}$, yielding $M'' \evalB M'''$ in \LF, where
$\ET{\Gamma'}{\Sigma'}(M) = M''$, and
$\ET{\Gamma'}{\Sigma'}(M')=M'''$, and
$\ET{\Gamma'}{\Sigma'}(\sigma)=\sigma'$. Since in \LF\ subject
reduction holds, then there is a derivation
$\D_3:\Gamma'\VDASH_{\Sigma'} M''':\sigma'$. Thus, we can ``decode''
$\D_3$ in \LLFP\ (via $\delta$), yielding a derivation
$\D'_3:\Gamma\VDASHS
\DT{\Sigma'}{\Gamma'}(M'''):\DT{\Sigma'}{\Gamma'}(\sigma')$,
\ie, a derivation $\D'_3:\Gamma\VDASHS M':\sigma$.
\end{thm}

\subsection{Other properties}
In a similar way
we can prove also other standard metatheoretic results:

\begin{prop}[Weakening and permutation]\label{pem:weak:s} If
  predicates are closed under signature/context weakening and
  permutation, then:
  \begin{enumerate}
  \item If $\Sigma$ and $\Omega$ are valid signatures, and every
    declaration occurring in $\Sigma$ also occurs in $\Omega$, then
    $\Gamma \VDASHS \alpha$ implies $\Gamma \VDASHO \alpha$.
  \item If $\Gamma$ and $\Delta$ are valid contexts \wrt\ the
    signature $\Sigma$, and every declaration occurring in $\Gamma$
    also occurs in $\Delta$, then $\Gamma \VDASHS \alpha$ implies
    $\Delta \VDASHS \alpha$.
\end{enumerate}
\end{prop}

\begin{prop}[Transitivity]
  \label{prp:trans} If predicates are closed under signature/context
  weakening and permutation and under substitution, then: if
  $\Gamma, x \of \sigma, \Gamma' \VDASHS \alpha$, and
  $\Gamma \VDASHS N : \sigma$, then
  $\Gamma, \Gamma'[N/x] \VDASHS \alpha[N/x]$.
\end{prop}

As for the so-called \emph{subderivation properties}, we need to be more
careful, as is shown in the following section.  The issue of
decidability for \LLFP\ can be addressed as that for \LFP\
in~\cite{honsell:hal-00906391}.

\subsection{Expressivity}\label{subsec:expr}
We recall that a system $\mathcal S'$ is a conservative extension of
$\mathcal S$ if the language of $\mathcal S$ is included in that of
$\mathcal S'$, and moreover for all judgements $\mathcal J$, in the
language of $\mathcal S$, then $\mathcal J$ is provable in
$\mathcal S'$ if and only if $\mathcal J$ is provable in $\mathcal S$.

\begin{thm} \LLFP\ is a conservative extension of \LF.
  \Proof \emph{(sketch)} The \emph{if} part is trivial. For the
  \emph{only if} part, consider a derivation in \LLFP\ and drop all
  locks/un\-locks (\ie\ \emph{release} the terms and types originally
  locked). This pruned derivation is a legal derivation in standard
  \LF.
\end{thm}
Notice that the above result holds independently of the particular
nature or any property of the external oracles that we may
\emph{invoke} during the proof development (in \LLFP), \eg\
decidability or recursive enumerability of $\P$.

Instead, \LLFP\ is \emph{not} a conservative extension of \LFP, since
the new typing rule allows us to derive more judgements with
unlocked-terms even if the predicate does not hold \eg\
$$
\infer[(O{\cdot}Guarded{\cdot}Unlock)] {\Gamma\VDASHS {\Lock \P S
    \sigma
    {x[\Unlock{\P}{S}{\sigma}{N}/x]}}:\Lock{\P}{S}{\sigma}{\tau[\Unlock{\P}{S}{\sigma}{N}/x]}}
{\Gamma,x\of\tau\VDASHS \Lock{\P}{S}{\sigma}{x} :
  \Lock{\P}{S}{\sigma}{\tau} & \Gamma\VDASHS N :
  {\Lock{\P}{S}{\sigma}{\tau}} & S\eqBL S & \sigma \eqBL \sigma}
$$
\noindent Then, since $x$ does not occur free in $\tau$,
$\Lock{\P}{S}{\sigma}{\tau[\Unlock{\P}{S}{\sigma}{N}/x]}\equiv
\Lock{\P}{S}{\sigma}{\tau}$
and we get
$\Gamma\VDASHS \Lock \P S \sigma {\Unlock{\P}{S}{\sigma}{N}} :
\Lock{\P}{S}{\sigma}{\tau}$.
This can be considered as the analogue of an $\eta$-expansion of
$\Gamma\VDASHS N : {\Lock \P S \sigma \tau}$ and it cannot be carried
out in plain \LFP\ if $\P(\Gamma\VDASHS S : \sigma)$ does not hold.

However, as we noticed at the end of Section~\ref{sec:lfp}, in the
Guarded Unlock Rules we require that the subject of the first premise
be necessarily externally locked. This fact, even in the presence of
locked variables in the typing context, avoids to derive unlocked
terms at top level. Indeed, we have the following:



%
%
\begin{thm}[Soundness of unlock]\label{th:unlock_soundness}
  If $\Gamma\VDASHS \Unlock{\P}{N}{\sigma}{M} : \tau$ is derived in
  \LLFP\
  then ${\P}(\Gamma\VDASHS N :\sigma)$ is true.

\Proof The proof can be carried out by a straightforward induction on the
  derivation of $\Gamma\VDASHS \Unlock{\P}{N}{\sigma}{M} : \tau$. So
  doing, we immediately restrict the possibilities for the last rule
  used in the derivation to $(O{\cdot}Top{\cdot}Unlock)$ and to
  $(O{\cdot}Conv)$ which affect only on the type of the judgment.
\end{thm}


Nevertheless, we have to phrase the so-called subderivation
properties carefully. Indeed, in \LLFP, given a derivation of
$\Gamma \VDASHS \alpha$ and a subterm $N$ occurring in the subject of
this judgement, we cannot prove that there always exists a derivation
of the form $\Gamma\VDASHS N:\tau$ (for a suitable $\tau$). Consider,
for instance, the previous example concerning the derivation of
$\Gamma\VDASHS \Lock{\P}{S}{\sigma}{\Unlock{\P}{S}{\sigma}{N}}:
\Lock{\P}{S}{\sigma}{\tau}$.
Clearly, if $\P(\Gamma\VDASHS S : \sigma)$ does not hold, then we
cannot derive any judgement whose subject and predicate are
$\Unlock{\P}{S}{\sigma}{N} :  \tau$.

Hence we have to restate point 6 of Proposition 3.11 (Subderivation,
part 1) of~\cite{honsell:hal-00906391} in a way similar to what we did
in~\cite{llfp-mfcs2014}.

\begin{prop}[Subderivation, part 1, point 6]\label{prp:subder:1}
  Given a derivation $\D : \Gamma \VDASHS \alpha$, and a subterm $N$
  occurring in the subject of this judgement, we have that either
  there exists a subderivation of a judgement having $N$ as a subject,
  or there exists a derivation of a judgment having as subject
  $\Lock{\P}{S}{\sigma}{N}$ (for suitable $\P$, $S$, $\sigma$).

  \Proof The proof is carried out by induction on the derivation of
  $\Gamma \VDASHS \alpha$.

\end{prop}

\section{Case Studies}\label{sec:cases}
In this section we discuss encodings of logics in \LLFP. Of course,
all encodings given in~\cite{honsell:hal-00906391} for \LFP, carry
over immediately to the setting of \LLFP, because the latter is a
language extension of the former. So here, we do not present encodings
for \emph{modal} and \emph{ordered linear logic}. However, the
possibility of using guarded unlocks, \ie\ the full power of the monad
destructor, allows for significant simplifications in several of the
encodings of logical systems given in \LFP. We illustrate this point
discussing call-by-value $\lambda_{\sf v}$-calculus, which greatly
benefits from the possibility of applying functions to
locked-arguments, and Hoare's Logic, which combines various kinds of
syntactical and semantical locks in its rules. We do not discuss
adequacy of these encodings since it is a trivial variant of the one
presented in~\cite{honsell:hal-00906391}.  One of the crucial problems
in designing restricted logical systems is to enforce
\emph{incrementally}, \emph{locally}, in rule application
\emph{global} constraints on derivations.  \LLFP\ can be very useful
in this respect since it allows one to focus precisely on the shape of
the well-behaved predicate. A classical case in point is \emph{Fitch
  Naive Set Theory} as formalized \eg\ in Prawitz \cite{prawitz}. The
global constraint is that derivations be normalizable. This means that
the elimination rules must not generate non-normalizable
derivations. In order to enforce this using a Lock-type, we need to
introduce a well-behaved predicate, \ie\ closed under
substitution. This is easy if the proof is closed, \ie\ it does not
involve assumptions, otherwise we must make sure that no future
instantiations are made on the variables corresponding to the
assumptions in the proof. To achieve this we introduce two kinds of
judgements the \emph{apodictic judgements}, \ie\ those which are
actually involved in the proofs and the \emph{generic}
ones. Unspecific judgements appear only in assumptions and in order to
be used must be ``demoted'' to an \emph{apodictic} judgement. What
happens in terms within the scope of the demoting operator does not
matter for the validity of the predicate. Thus variables witnessing
generic assumptions, even if replaced, behave as constants while
variables witnessing an apodictic judgements can be freely
substituted. The local constraint in the elimination rules accesses
the input proof-terms, checks that the combination can be normalized
and furthermore that all free variables of judgement type are generic,
see Subsection~\ref{Fitch}.

\subsection{Call-by-value $\lambda_{\sf v}$-calculus}\label{cbv-sec}
\noindent We encode, using \textit{Higher Order Abstract Syntax}
(HOAS), the syntax of untyped $\lambda$-calculus:
$M,N ::= x\ |\ M\ N\ |\ \lambda x.M$ as
in~\cite{honsell:hal-00906391}, where natural numbers (through the
constructor \texttt{free}) are used to represent free variables, while
bound variables are rendered as metavariables of \LLFP\ of type
\texttt{term}:

\begin{defi}[\LLFP\ signature $\Sigma_\lambda$ for untyped
  $\lambda$-calculus]\label{signature-syntax-ulc}
\noindent
\begin{small}
\begin{alltt}
term : Type                      nat  :  Type            \(\hfill\)  O : nat
   S :  nat -> nat               free :   nat -> term
 app : term -> term -> term      lam  : (term -> term) -> term
\end{alltt}
\end{small}
\end{defi}

\begin{defi}[Call-by-value reduction strategy]\label{def:cbv}
  The call-by-value evaluation strategy is given by:
  \begin{small}
    $$
    \begin{array}{l@{\quad\quad}l}
      \infer[\sf (refl)]
      {\VDASH_{\sf v} M=M}
      {}
      &
      \infer[\sf (symm)]
      {\VDASH_{\sf v} M=N}
      {\VDASH_{\sf v} N=M}
      \\[3mm]
      \infer[\sf (trans)]
      {\VDASH_{\sf v} M=P}
      {\VDASH_{\sf v} M=N \quad \VDASH_{\sf v} N=P}
      \quad
      &
      \infer[\sf (app)]
      {\VDASH_{\sf v} M \at M'=N \at N'}
      {\VDASH_{\sf v} M=N \quad \VDASH_{\sf v} M'=N'}
      \\[3mm]
      \infer[\sf (\beta_v)]
      {\VDASH_{\sf v} (\lambda x.M) \at v=M[v/x]}
      {\mbox{$v$ is a value}}
      &
      \infer[\sf (\xi_v)]
      {\VDASH_{\sf v} \lambda x.M=\lambda x.N}
      {\VDASH_{\sf v} M=N}
    \end{array}
    $$
  \end{small}
  where values are either variables,  constants, or
  abstractions.
\end{defi}

The new typing rule $(O{\cdot}Guarded{\cdot}Unlock)$ of \LLFP, allows
to encode naturally the system as follows.

\begin{defi}[\LLFP\ signature $\Sigma_{\sf v}$ for
  $\lambda_{\sf v}$-calculus]\label{signature-reduction-ml}
  We extend the signature of Definition \ref{signature-syntax-ulc} as
  follows:
  \begin{small}
\begin{alltt}
   eq : term->term->Type
 refl : \(\Pi\)M:term.(eq M M)
 symm : \(\Pi\)M:term.\(\Pi\)N:term.(eq N M)->(eq M N)
trans : \(\Pi\)M,N,P:term.    \,\,(eq M N)->(eq N P)\,\,\,->(eq M P)
eq\_app \!\!: \(\Pi\)M,N,M'\!,N':term.\,\,\,(eq M N)->(eq M'N')->(eq (app M M')(app N N'))
betav : \(\Pi\)M:(term->term).   \(\Pi\)N:term.\(\Lock{\mbox{\scriptsize\emph{Val}}}{\mathtt{N}}{\mathtt{term}}{\mathtt{(eq (app (lam M) N)(M N))}}\)
 csiv : \(\Pi\)M,N:(term->term).(\(\Pi\)x:term.\(\Lock{\mbox{\scriptsize\emph{Val}}}{\mathtt{x}}{\mathtt{term}}{\mbox{\tt(eq (M x)(N x))}}\))->(eq (lam M)(lam N))
\end{alltt}
\end{small}

\noindent where the predicate {\sf {Val}} is defined as
follows:\\
\noindent\phantom{--}-- {\tt \emph{Val}}$(\Gamma\VDASHS$ {\tt N :
  term}$)$ holds iff either {\tt N} is an abstraction or a constant
(\ie\ a term of the shape {\tt{(free i)}}).
\end{defi}
Notice the neat improvement \wrt\ to the encoding of \LFP, given
in~\cite{honsell:hal-00906391}, as far as the rule \texttt{csiv}. The
encoding of the rule $\sf \xi_v$ is problematic if bound variables are
encoded using metavariables, because the predicate {\tt Val} appearing
in the lock cannot mention explicitly variables, for it to be
\emph{well-behaved}. In~\cite{honsell:hal-00906391}, since we could
not apply the rules unless we had explicitly eliminated the {\tt
  Val}-lock, in order to overcome the difficulty we had to make a
detour using constants. In \LLFP, on the other hand, we can apply the
rules ``under {\tt Val}'', so to speak, and postpone the proof of the
{\tt Val}-checks till the very end, and then rather than checking {\tt
  Val} we can get rid of the lock altogether, since the bound variable
of the rule \texttt{csiv}, is assumed to be locked. Notice that this
phrasing of the rule \texttt{csiv} amounts precisely to the fact that
in $\lambda_{\sf v}$ variables range over values. As a concrete
example of all this, we show how to derive the equation
$\lambda x.z \at ((\lambda y.y) \at x) = \lambda x.z \at x$. Using
``pencil and paper'' we would proceed as follows:
$$
\infer[\sf (\xi_v)]{\VDASH_{\sf v}\lambda x.z \at ((\lambda y.y) \at
  x)=\lambda x.z \at x} { \infer[\sf (app)]{\VDASH_{\sf v} z \at
    ((\lambda y.y) \at x)=z \at x} {\infer[\sf
    (refl)]{\VDASH_{\sf v} z=z} {-} \qquad \infer[\sf (\beta_v)]{(\lambda
      y.y) \at x=y[x/y]} {\mbox{$x$ is a value}} } }
$$
\noindent Similarly, in \LLFP, we can derive
$\mathtt{z{:}term}\VDASHS\mathtt{(refl\ z):(eq\ z\ z)}$ and
$$
\Gamma,\mathtt{x{:}term}\VDASHS\mathtt{(betav\at (\lambda
  y{:}term.y)\at x)}:\Lock{\mbox{\scriptsize {\tt Val }
  }}{\mathtt{x}}{\mathtt{term}}{\mathtt{(eq\at (app\at (lam\at \lambda
    y{:}term.y)\at x)\at ((\lambda y{:}term.y) \at x))}}.
$$
\!\!This far, in old \LFP, we would be blocked if we could not prove
that {\tt Val}$(\Gamma,{\tt x{:}term}$ $\VDASHS {\tt x : term})$
holds, since \texttt{eq\_app} cannot accept an argument with a
locked-type. However, in \LLFP, we can apply the
$(O{\cdot}Guarded{\cdot}Unlock)$ rule obtaining the following proof
term (from the typing environment $\Gamma,{\tt x{:}term, z{:}term}$):
$$
\Lock {\tt Val} {\mathtt{x}} {\mathtt{term}} {\mathtt{(eq\_app\ z\ z\
    (app\ (lam\ \lambda y{:}term.y)\ x)\ x\ (refl\ z)\
    \Unlock{\mbox{\scriptsize
        {\tt Val}}}{\mathtt{x}}{\mathtt{term}}{\mathtt{(betav\
        (\lambda y{:}term.y)\ x)}})}}
$$
of type
$\Lock{\mbox{\scriptsize{\tt
      Val}}}{\mathtt{x}}{\mathtt{term}}{\mathtt{(eq\ (app\ z\ (app\
    (lam\ \lambda y{:}term.y)\ x))\ (app\ z\ x))}}$.
And abstracting $\tt x$, a direct application of \texttt{csiv} yields
the result.

\subsection{Imp with Hoare Logic}
\noindent An area of Logic which can greatly benefit from the new
system \LLFP\ is \emph{program logics}, because of the many
syntactical checks which occur in these systems. To illustrate this
fact, we consider a very simple imperative language Imp, whose syntax
is:
\begin{center}
  $\begin{array}{lcl@{\qquad\qquad}r}
     p & ::= & skip\ |\ x := expr\ |\ p;p\ | & \mbox{\textrm{\hfill{null $|$ assignment $|$ sequence}}}\\
       & & i\!f\ cond\ then\ p\ else\ p\ |\ while\ cond\ \{p\}& \mbox{\hfill{cond  $|$ while}}
  \end{array}$
\end{center}
\noindent Other primitive notions of Imp are variables, both integer
and \emph{identifier}, and expressions. Identifiers denote
locations. \noindent For the sake of simplicity, we assume only
integers (represented by type {\tt int}) as possible values for
identifiers. In this section, we follow as closely as possible the
HOAS encoding, originally proposed in \cite{AHMP-92}, in order to
illustrate the features and possible advantages of using \LLFP\ \wrt\
\LF. The main difference with that approach is that here we encode
\emph{concrete} identifiers by constants of type \texttt{var}, an {\tt
  int}-like type, of course different from {\tt int} itself, so as to
avoid confusion with possible values of locations.

\begin{defi}[\LLFP\ signature $\Sigma_{\sf Imp}$ for Imp]
\label{signature-syntax-imp}\hfill
\begin{alltt}
 int : Type                           bool : Type	
 var : Type                        and,imp : bool -> bool -> bool
bang :  var -> int                  0,1,-1 :  int
   = :  int -> int -> bool               + :  int ->  int -> int
 not : bool -> bool                 forall : (int -> bool) -> bool
\end{alltt}
\end{defi}
\noindent Since variables of type \texttt{int} may be bound in
expressions (by means of the \texttt{forall} constructor), we define
explicitly the encoding function $\epsilon^{\tt{exp}}_{\mathcal{X}}$
mapping expressions with free variables of type {\tt int} in
${\mathcal{X}}$ of the source language Imp into the corresponding
terms of \LLFP:

\begin{center}
$\begin{array}{rcl@{\quad}rcl}
\epsilon^{\tt{exp}}_{\mathcal X}(0)                    & \eqdef & \mbox{\tt 0} &
\epsilon^{\tt{exp}}_{\mathcal X}({+/-}1)            & \eqdef &  \mbox{\tt +/-1}\\[3mm]
\epsilon^{\tt{exp}}_{\mathcal X}(x)                   & \eqdef & {\left\{\begin{array}{l@{~}l}{\tt x} &
 \mbox{if }  x\in{\mathcal X}\\
{\tt (bang~x)} &\mbox{if } x\not\in{\mathcal
     X}\end{array}\right.}   &     &\\[5mm]
\epsilon^{\tt{exp}}_{\mathcal{X}}(n+m)             & \eqdef &  (\mbox{\tt +}~\epsilon^{\tt{exp}}_
{\mathcal{X}}(n)\, \epsilon^{\tt{exp}}_ {\mathcal{X}}(m)) &
\epsilon^{\tt{exp}}_{\mathcal{X}}(n=m)             & \eqdef &  (\mbox{\tt
  =}~\epsilon^{\tt{exp}}_{\mathcal{X}}(n)\, \epsilon^{\tt{exp}}_{\mathcal{X}}(m))\\[5mm]
\epsilon^{\tt{exp}}_{\mathcal{X}}(\lnot e)          & \eqdef & (\mbox{\tt
  not}~\epsilon^{\tt{exp}}_{\mathcal{X}}(e)) &
\epsilon^{\tt{exp}}_{\mathcal{X}}(e\wedge e')    & \eqdef &  (\mbox{\tt
  and}~\epsilon^{\tt{exp}}_{\mathcal{X}}(e)\, \epsilon^{\tt{exp}}_{\mathcal{X}}(e'))\\[3mm]
\epsilon^{\tt{exp}}_{\mathcal{X}}(e\supseteq e') & \eqdef & (\mbox{\tt
  imp}~\epsilon^{\tt{exp}}_{\mathcal{X}}(e)\, \epsilon^{\tt{exp}}_{\mathcal{X}}(e')) &
\epsilon^{\tt{exp}}_{\mathcal{X}}(\forall x.\phi) & \eqdef &
(\mbox{\tt forall}~\lambda{\tt{x}}\of {\tt int}.\epsilon^{\tt{exp}}_{{\mathcal{X}}\cup\{x\}}(\phi))
\end{array}$
\end{center}

\noindent where \texttt{x} in \texttt{(bang x)} denotes the encoding
of the concrete memory location (\ie, a constant of type \texttt{var})
representing the (free) source language identifier $x$; the other case
represents the free variable $x$ rendered as a \LLFP\ metavariable
\texttt{x} of type \texttt{int} in HOAS style. The syntax of
imperative programs is defined as follows:

\begin{defi}[\LLFP\ signature $\Sigma_{\sf Imp}$ for Imp with command]
  \label{signature-syntax-cmd}\hfill\\
We extend the signature of Definition \ref{signature-syntax-imp} as
follows:

\begin{alltt}
  prog : Type
 Iskip : prog
  Iseq : prog -> prog -> prog
  Iset :  var ->  int -> prog
   Iif : \(\Pi\)e:bool.prog -> prog ->\(\Lock{\scriptsize\tt{QF}}{\scriptsize\tt{e}}{\scriptsize\tt{bool}}{\tt{prog}}\)
Iwhile : \(\Pi\)e:bool.prog -> \(\Lock{{\scriptsize\tt{QF}}}{\scriptsize\tt{e}}{\scriptsize\tt{bool}}{\tt{prog}}\)
\end{alltt}
\end{defi}

\noindent where the predicate {\tt QF}($\Gamma\VDASHSIMP$ {\tt
  e:bool}) holds iff the formula {\tt e} is \emph{closed} and
\emph{quantifier free}, \ie, it does not contain the {\tt forall}
constructor.  We can look at {\tt QF} as a ``good formation''
predicate, ruling out \emph{bad programs with invalid boolean
  expressions} by means of stuck terms.

The encoding function $\epsilon^{\tt prog}_{\mathcal{X}}$ mapping
programs with free variables in ${\mathcal{X}}$ of the source language
Imp into the corresponding terms of \LLFP\ is defined as follows:
\begin{alltt}
\(\epsilon\sp{\tt{prog}}\sb{\mathcal X}(skip)\)              \:\(\eqdef\) Iskip

\(\epsilon\sp{\tt{prog}}\sb{\mathcal X}(x:=e)\)             \!\(\eqdef\) (Iset x \(\epsilon\sp{\tt{exp}}\sb{\mathcal X}(e)\))

\(\epsilon\sp{\tt{prog}}\sb{\mathcal{X}}(p ; p')\)            \;\(\eqdef\) (Iseq \(\epsilon\sp{\tt{prog}}\sb{\mathcal{X}}(p)\) \(\epsilon\sp{\tt{prog}}\sb{\mathcal{X}}(p')\))

\(\epsilon\sp{\tt{prog}}\sb{\mathcal{X}}(\mathit{if} e then p else p')\) \(\eqdef\) \(\Unlock{\mbox{\scriptsize\tt{QF}}}{\epsilon\sp{\tt{exp}}\sb{\mathcal{X}}(e)}{\mbox{\scriptsize\tt{bool}}}{(\mathtt{Iif} \epsilon\sp{\tt{exp}}\sb{\mathcal{X}}(e) \epsilon\sp{\tt{prog}}\sb{\mathcal{X}}(p) \epsilon\sp{\tt{prog}}\sb{\mathcal{X}}(p'))}\) \mbox{\textrm{(*)}}

\(\epsilon\sp{\tt{prog}}\sb{\mathcal{X}}(while e \{p\})\)       \;\(\eqdef\) \(\Unlock{\mbox{\scriptsize\tt{QF}}}{\epsilon\sp{\tt{exp}}\sb{\mathcal{X}}(e)}{\mbox{\scriptsize\tt{bool}}}{(\mathtt{Iwhile} \epsilon\sp{\tt{exp}}\sb{\mathcal{X}}(e) \epsilon\sp{\tt{prog}}\sb{\mathcal{X}}(p))}\) \mbox{\textrm{(*)}}
\end{alltt}
(*) if $e$ is a quantifier-free formula.  However, in the last two
clauses, the terms on the right hand side cannot be directly expressed
in general form (\ie, for all expressions $e$) because if {\tt
  QF}($\Gamma\VDASHSIMP\epsilon\sp{\tt{exp}}\sb{\mathcal{X}}(e)$ {\tt
  : bool}) does not hold, we cannot use the unlock operator. Thus we
could be left with two terms of type
$\Lock{\mbox{\scriptsize\tt{QF}}}{\epsilon\sp{\tt{exp}}\sb{\mathcal{X}}(e)}{\tt{bool}}{\tt{prog}}$,
instead of type \texttt{prog}. This is precisely the limit of the
\LFP\ encoding in~\cite{honsell:hal-00906391}. Since a $\U$-term can
only be introduced if the corresponding predicate holds, when we
represent rules of Hoare Logic we are forced to consider only legal
terms, and this ultimately amounts to restricting explicitly the
object language in a way such that {\tt QF} always returns true.

In \LLFP, instead, we can use naturally the following signature for
representing Hoare's Logic, without assuming anything about the object
language terms (given the predicate \texttt{true : bool -> Type} such
that \texttt{(true e)} holds iff \texttt{e} is true):

\begin{defi}[\LLFP\ signature $\Sigma_{\sf HL}$ for Hoare Logics]
  \label{signature-syntax-hoare}\hfill

\begin{alltt}
        args : Type
       <_,_> :  var -> (int -> bool) -> args
       hoare : bool -> prog -> bool  -> Type\(\down{1}\)
 hoare_Iskip : \(\Pi\)e:bool.(hoare e Iskip e)\(\down{1}\)
  hoare_Iset : \(\Pi\)t:int.\(\Pi\)x:var.\(\Pi\)e:int -> bool.
  \hfill \(\ELock{\mbox{\scriptsize{P\(\sp{\tt{set}}\)}}}{}{\mbox{\scriptsize\tt{\(\langle\)x,e\(\rangle\),args}}}{\mbox{\tt(hoare (e t) (Iset x t) (e (bang x))}}\)\(\down{1}\)
  hoare_Iseq : \(\Pi\)e,e',e'':bool.\(\Pi\)p,p':prog.(hoare e  p  e')  ->
                                          (hoare e' p' e'') ->
                                          (hoare e (Iseq p p') e'')\(\down{1}\)
   hoare_Iif : \(\Pi\)e,e',b:bool.\(\Pi\)p,p':prog.(hoare (b and e) p e') ->
                                        (hoare ((not b) and e) p' e') ->
                                        \(\Lock{\mbox{\tt\scriptsize{QF}}}{\tt{b}}{\mbox{\scriptsize\tt{bool}}}{\mathtt{(hoare e }\Unlock{\mbox{\tt\scriptsize{QF}}}{\tt{b}}{\mbox{\scriptsize\tt{bool}}}{\tt(Iif b p p')}\mathtt{ e')}}\)\(\down{1}\)
hoare_Iwhile : \(\Pi\)e,b:bool.\(\Pi\)p:prog.(hoare (e and b) p e) ->
   \hfill \(\Lock{\mbox{\tt\scriptsize{QF}}}{\tt{b}}{\mbox{\scriptsize\tt{bool}}}{\mathtt{(hoare e }\Unlock{\mbox{\tt\scriptsize{QF}}}{\tt{b}}{\mbox{\scriptsize\tt{bool}}}{\tt(Iwhile b p)}\mathtt{ ((not b) and e))}}\)\(\down{1}\)
 hoare_Icons : \(\Pi\)e,e',f,f':bool.\(\Pi\)p:prog.(true (imp e' e)) ->
                                        (hoare e  p f)    ->
                                        (true (imp f f')) ->
                                        (hoare e' p f')
\end{alltt}
\end{defi}
\noindent where {\tt P}$^{\tt set}$($\Gamma\VDASHSHOARE \langle${\tt
  x},{\tt e}$\rangle$ : {\tt args}) holds iff {\tt e} is
closed\footnote{Otherwise, the predicate {\tt \emph{P}} would not be
  well-behaved, see Definition~\ref{def:wbred}.} and the location
(\ie, constant) {\tt x} does not occur in {\tt e}. Such requirements
amount to formalizing that no assignment made to the location denoted
by {\tt x} affects the meaning or value of {\tt e}
(\emph{non-interference} property). The intuitive idea here is that
\begin{verse}
  if {\tt e=}$\epsilon^{\tt{exp}}_{\mathcal X}(E)$, and {\tt
    p=}$\epsilon^{\tt{prog}}_{\mathcal{X}}(P)$, and {\tt
    e'=}$\epsilon^{\tt{exp}}_{\mathcal{X}}(E')$ hold,\\ then \texttt{(hoare
    e p e')} holds iff the Hoare's triple $\{E\}P\{E'\}$ holds.
\end{verse}
The advantage \wrt\ previous encodings (see, \eg, \cite{AHMP-92}), is
that in \LLFP\ we can delegate to the external predicates {\tt QF} and
{\tt P}$^{\tt set}$ all the complicated and tedious checks concerning
\emph{non-interference} of variables and good formation clauses for
guards in the conditional and looping statements. Thus, the use of
lock-types, which are subject to the verification of such conditions,
allows to legally derive $\Gamma\VDASHSHOARE$ {\tt m} : {\tt (hoare e
  p e')} only according to the Hoare semantics.

Moreover, the $(O{\cdot}Guarded{\cdot}Unlock)$ rule allows also to
``postpone'' the verification that {\tt QF}($\Gamma\VDASHS$ {\tt
  e : bool}) holds (\ie, that the formula {\tt e} is \emph{quantifier
  free}).

\subsection{Fitch Set Theory \ala\ Prawitz}
\label{Fitch}
In this section we present the encoding of a logical system of
remarkable logical, as well as historical, significance, namely the
system of consistent \emph{Naive} Set Theory, {\tt{F}} due to Fitch
\cite{fitch}. This system, was first presented in Natural Deduction
style by Prawitz \cite{prawitz}. Naive Set Theory, being inconsistent,
in order to prevent the derivation of inconsistencies from the
unrestricted \emph{abstraction} rule, in the system {\tt{F}} only
normalizable \emph{deductions} are allowed. Of course this side
condition in the rule is extremely difficult to capture using
traditional tools.

In the present context, instead, we can put to use the machinery of
\LLFP\ to provide an appropriate encoding of {\tt{F}} where the
\emph{global} normalization constraint is enforced \emph{locally} by
checking the proof-object. This system is a beautiful example for
illustrating the \emph{bag of tricks} of \LLFP. Checking that a proof
term is normalizable would be the obvious predicate to use in the
corresponding lock type, but this would not be a well-behaved
predicate if free variables, \ie\ assumptions, are not sterilized,
because predicates need to be well-behaved. To this end we introduce a
distinction between \emph{generic} judgements, which cannot be
directly utilized in arguments, but which can be assumed and
\emph{apodictic} judgements, which are directly involved in proof
rules. In order to make use of generic judgements, one has to
downgrade them to an apodictic one. This is achieved by a suitable
coercion function.\newpage

\begin{defi}[\LLFP\ signature $\Sigma_{\sf FPST}$ for Fitch Prawitz
  Set Theory]\label{signature-fitch}
  The  following constants are introduced:
\begin{alltt}
o   : Type
\(\iota\)   \,: Type
T   : o -> Type
V   : o -> Type
lam : (\(\iota\) -> o)-> \(\iota\)
\(\forall\)   : (\(\iota\) -> o)-> o
\(\epsilon\)   \,: \(\iota\) -> (\(\iota\) -> o ) -> o
\(\supset \)  \!: o -> o -> o
\(\delta\)        \!: \(\Pi\)A:o.  (V(A) -> T(A))
\(\mathtt{\supset_{intro}}\) : \(\Pi\)A,B:o.(V(A) -> T(B)) -> (T(A \(\supset\)B))
\(\mathtt{\supset_{elim}}\)  : \(\Pi\)A,B:o.\(\Pi\)x:T(A).\(\Pi\)y:T(A\(\supset\)B) -> \({\Lock{\scriptsize\tt{Fitch}}{\scriptsize{\mathtt{\langle{x},y\rangle}}}{\scriptsize\mathtt{{T(A)}\times\mathtt{V(A)->T(B)}}}{\mathtt{T(B)}}}\)
\end{alltt}
  \noindent where {\tt o} is the type of propositions, $\supset$ is
  the syntactic constructor for propositions together with the
  ``membership" predicate $\epsilon$ and {\tt lam} is the
  ``abstraction" operator for building ``sets", $\mathtt{T}$ is the
  apodictic judgement, while $\mathtt{V}$ s the generic judgement, and
  $\mathtt{<x,y>}$ denotes the encoding of pairs, whose type is
  denoted by $\mathtt{\sigma X \tau}$, \eg\
  $$\mathtt{\lambda u\of\sigma \rightarrow \tau \rightarrow \rho.\ u\
    x\ y : (\sigma \rightarrow \tau} \mathtt{\rightarrow
    \rho)\rightarrow \rho}$$
  The predicate in the lock is defined as follows:

  \begin{center}
    {\tt Fitch}($\Gamma\vdash_{\sf FPST} \mathtt{<x,y>\ : T(A)X
      V(A)\rightarrow T(B)}$)
  \end{center}

  \noindent it holds iff the proof derived by combining $\mathtt{x}$
  and $\mathtt{y}$ is normalizable and all occurrences of free
  variables of judgement type occur within the scope of a $\delta$.

\end{defi}

For lack of space we do not spell out the rules concerning the other
logical operators, because they are all straightforward provided we
use only the apodictic judgement $\mathtt{T(\cdot)}$. But a few
remarks are mandatory. The notion of \emph{normalizable proof} is the
standard notion of normal proof used in natural deduction. The
predicate {\tt Fitch} is well-behaved because free judgement variables
cannot be replaced by any sensible object.  Adequacy for this
signature can be achieved in the general formulation of
\cite{honsell:hal-00906391}, namely:

\begin{thm}[Adequacy for Fitch-Prawitz Naive Set Theory]
  $A_1\ldots A_n \vdash_{\sf FPST} A $ iff there exists a normalizable
  $\mathtt{M}$ such that
  $\mathtt{x_1\of V(A_1),\ldots,x_n\of V(A_n)} \vdash_{\sf FPST}
  \mathtt{M :T(A)}$.
\end{thm}

\section{Concluding remarks: from Predicates to Functions and
  beyond}\label{sec:conclusion}
We have shown how to extend \LF\ with a class of monads which capture
the effect of delegating to an external oracle the task of providing
part of the necessary evidence for establishing a judgement. Thus we
have extended with an additional clause the \LF\ paradigm for encoding
a logic, namely: \emph{external evidence as monads}. This class of
monads is very well-behaved and so it permits to simplify the
equational theory of the system. In principle we could have used the
$let_T$ destructor, together with its equational theory as in Moggi's
general approach \cite{Moggi-Computationallambda}, but we think that
our approach greatly simplifies the theory, since it does away with
permutative reductions.

The technique for proving the metatheoretic properties of \LLFP\ that
we used in this paper is rather powerful and novel. It generalizes the
technique that was traditionally used to prove normalization
properties for systems with dependent types, by reducing such
languages to the corresponding dependency-less systems, see
\cite{Bar-92}. In this paper we have actually managed to reduce the
whole proof system \LLFP\ to \LF\, as well as to translate it back,
thereby transferring nearly all metatheoretic properties of \LF\ to
our new system.

We have presented \LLFP\ in the standard style, but as future work we
want to move to the \emph{canonical} style of
\cite{HarperLicata-jfp-07} in vision of a future prototype
implementation.

In this paper we consider the verification of predicates in locks as
purely \emph{atomic actions}, \ie\ each predicate \perse. But of
course predicates have a logical structure which can be reflected onto
locks. \Eg\ we can consistently extend \LLFP\ by assuming that locks
\emph{commute, combine}, and \emph{entail}, \ie\ that the following
types are inhabited:
$$\begin{array}{l}
    \Lock{\P}{x}{\sigma}{\tau} \rightarrow \Lock{\Q}{x}{\sigma}{\tau} \mbox{  if}\\[2mm]
    \P(\Gamma \VDASHS x : \sigma ) \rightarrow \Q(\Gamma \VDASHS x :
    \sigma) \mbox{, and}\\[2mm]
    \Lock{\P}{x}{\sigma}{\Lock{\Q}{x}{\sigma}{M}}\rightarrow \Lock{\P \&
                                                                         \Q}{x}{\sigma}{M} \mbox{, and}\\[2mm]
    \Lock{\P}{x}{\sigma}{\Lock{\Q}{y}{\tau}{M}}\rightarrow
    \Lock{\Q}{y}{\tau}{\Lock{\P}{x}{\sigma}{M}}.
\end{array}$$
We encoded call-by-value $\lambda$-calculus with Plotkin's classical
notion of \emph{value}. But the encoding remains the same, apart from
what is delegated to the lock, if we consider other notions of value
\eg\ \emph{closed normal forms} only for $K$-redexes
\cite{honsell1999semantical}. The example of Fitch's system suggests
further generalization, and illustrates how monads handle
side-conditions uniformly.

\medskip Yet, as a near future work, another interesting direction
will be to extend \LLFP\ to oracle-calls which produce an {\em
  output}. Namely, rather then just having external predicates which
check that a judgement satisfies a given property, we could give the
oracle a \emph{query} and let it provide the {\em witness}.  More
precisely the lock operator \(\mathcal{L}\)
could bind a particular variable $\mathtt{x}$ in $\mathtt{M}$ that
needs to be instantiated. The predicate in the lock would then become
a sort of query on $\mathtt{x}$, which could be fed to the oracle. If
successful, the unlock operator could provide then this witness. The
Unlock/Lock reduction would amount to replacing $\mathtt{x}$ by the
witness. Suitable \emph{compilation} and \emph{decompilation}
functions between \LF\ and the language of the oracle should allow for
the correct expression of the witness.

\bibliographystyle{abbrv}
\bibliography{LFP}

\begin{thebibliography}{10}

\bibitem{albert1968}
H.~Albert.
\newblock {\em Traktat \"uber kritische Vernunft}.
\newblock J.C.B. Mohr (Paul Siebeck), T\"ubingen, 1991.

\bibitem{alechina2001}
N.~Alechina, M.~Mendler, V.~De~Paiva, and E.~Ritter.
\newblock Categorical and kripke semantics for constructive s4 modal logic.
\newblock In {\em Computer Science Logic}, pages 292--307. Springer, 2001.

\bibitem{acerbi}
Archimede.
\newblock {\em {Metodo. Nel laboratorio di un genio}}.
\newblock Bollati Boringhieri, 2013.

\bibitem{AHMP-92}
A.~Avron, F.~Honsell, I.~Mason, and R.~Pollack.
\newblock {Using Typed Lambda Calculus to Implement Formal Systems on a
  Machine}.
\newblock {\em Journal of Automated Reasoning}, 9(3):309--354, 1992.

\bibitem{badiou}
A.~Badiou.
\newblock {\em {La relation {\'{e}}nigmatique entre philosophie et politique}}.
\newblock Germina, 2011.

\bibitem{Bar-book}
H.~Barendregt.
\newblock {\em Lambda Calculus: its Syntax and Semantics}.
\newblock North Holland, 1984.

\bibitem{Bar-92}
H.~Barendregt.
\newblock Lambda {C}alculi with {T}ypes.
\newblock In {\em Handbook of Logic in Computer Science}, volume~II, pages
  118--310. Oxford University Press, 1992.

\bibitem{bar02}
H.~Barendregt and E.~Barendsen.
\newblock Autarkic computations in formal proofs.
\newblock {\em Journal of Automated Reasoning}, 28:321--336, 2002.

\bibitem{BCKL-03}
G.~Barthe, H.~Cirstea, C.~Kirchner, and L.~Liquori.
\newblock Pure {P}attern {T}ype {S}ystems.
\newblock In {\em POPL'03}, pages 250--261. The ACM Press, 2003.

\bibitem{deep}
R.~Boulton, A.~Gordon, M.~Gordon, J.~Harrison, J.~Herbert, and J.~V. Tassel.
\newblock Experience with embedding hardware description languages in hol.
\newblock In {\em Theorem Provers in Circuit Design, TPCD}, pages 129--156.
  North-Holland, 1992.

\bibitem{carr}
L.~Carroll.
\newblock What the {T}ortoise {S}aid to {A}chilles.
\newblock {\em Mind}, 4:278--280, 1895.

\bibitem{LF-modulo}
D.~Cousineau and G.~Dowek.
\newblock Embedding pure type systems in the lambda-pi-calculus modulo.
\newblock In {\em Typed Lambda Calculi and Applications, TLCA}, volume 4583 of
  {\em Lecture Notes in Computer Science}, pages 102--117. Springer-Verlag,
  2007.

\bibitem{fairtlough1997propositional}
M.~Fairtlough and M.~Mendler.
\newblock Propositional lax logic.
\newblock {\em Information and Computation}, 137(1):1--33, 1997.

\bibitem{fairtlough2001abstraction}
M.~Fairtlough, M.~Mendler, and X.~Cheng.
\newblock Abstraction and refinement in higher order logic.
\newblock In {\em Theorem Proving in Higher Order Logics}, pages 201--216.
  Springer, 2001.

\bibitem{Fairtlough97first-orderlax}
M.~Fairtlough, M.~Mendler, and M.~Walton.
\newblock First-order lax logic as a framework for constraint logic
  programming.
\newblock Technical report, 1997.

\bibitem{fitch}
F.~B. Fitch.
\newblock {\em {Symbolic logic}}.
\newblock New York, 1952.

\bibitem{garg2008indexed}
D.~Garg and M.~C. Tschantz.
\newblock From indexed lax logic to intuitionistic logic.
\newblock Technical report, DTIC Document, 2008.

\bibitem{girard2011blind}
J.-Y. Girard.
\newblock {\em The Blind Spot: lectures on logic}.
\newblock European Mathematical Society, 2011.

\bibitem{HHP-92}
R.~Harper, F.~Honsell, and G.~Plotkin.
\newblock A {F}ramework for {D}efining {L}ogics.
\newblock {\em Journal of the ACM}, 40(1):143--184, 1993.
\newblock Preliminary version in Proc. of {LICS'87}.

\bibitem{HarperLicata-jfp-07}
R.~Harper and D.~Licata.
\newblock Mechanizing metatheory in a logical framework.
\newblock {\em J. Funct. Program.}, 17:613--673, 2007.

\bibitem{hirschkoff:bisimproofs}
D.~Hirschkoff.
\newblock Bisimulation proofs for the $\pi$-calculus in the {C}alculus of
  {C}onstructions.
\newblock In {\em Proc.~TPHOL'97}, number 1275 in LNCS. Springer-Verlag, 1997.

\bibitem{Honsell:2013:YFP:2503887.2503896}
F.~Honsell.
\newblock 25 years of formal proof cultures: Some problems, some philosophy,
  bright future.
\newblock In {\em Proceedings of the Eighth ACM SIGPLAN International Workshop
  on Logical Frameworks and Meta-languages: Theory and Practice}, LFMTP'13,
  pages 37--42, New York, NY, USA, 2013. ACM.

\bibitem{honsell1999semantical}
F.~Honsell and M.~Lenisa.
\newblock Semantical analysis of perpetual strategies in $\lambda$-calculus.
\newblock {\em Theoretical Computer Science}, 212(1):183--209, 1999.

\bibitem{HLL06}
F.~Honsell, M.~Lenisa, and L.~Liquori.
\newblock {A Framework for Defining Logical Frameworks}.
\newblock {\em Volume in Honor of G. Plotkin, ENTCS}, 172:399--436, 2007.

\bibitem{honsell:hal-00906391}
F.~Honsell, M.~Lenisa, L.~Liquori, P.~Maksimovic, and I.~Scagnetto.
\newblock {An Open Logical Framework}.
\newblock {\em Journal of Logic and Computation}, Oct. 2013.

\bibitem{HLLS08}
F.~Honsell, M.~Lenisa, L.~Liquori, and I.~Scagnetto.
\newblock A conditional logical framework.
\newblock In {\em LPAR'08}, volume 5330 of {\em LNCS}, pages 143--157.
  Springer-Verlag, 2008.

\bibitem{llfp-mfcs2014}
F.~Honsell, L.~Liquori, and I.~Scagnetto.
\newblock {\LaxLFP: Side Conditions and External Evidence as Monads}.
\newblock In {\em {Proc. of MFCS 2014 (39th International Symposium on
  Mathematical Foundations of Computer Science), Part I}}, volume 8634 of {\em
  Lecture Notes in Computer Science}, pages 327--339, Budapest, Hungary, August
  2014. Springer.

\bibitem{HMS-01}
F.~Honsell, M.~Miculan, and I.~Scagnetto.
\newblock {$\pi$-calculus in (Co)Inductive Type Theories}.
\newblock {\em Theoretical Computer Science}, 253(2):239--285, 2001.

\bibitem{mendler1991constrained}
M.~Mendler.
\newblock Constrained proofs: A logic for dealing with behavioural constraints
  in formal hardware verification.
\newblock In {\em Designing Correct Circuits}, pages 1--28. Springer-Verlag,
  1991.

\bibitem{moggi1988partial}
E.~Moggi.
\newblock {\em The partial lambda calculus}.
\newblock PhD thesis, University of Edinburgh. College of Science and
  Engineering. School of Informatics, 1988.

\bibitem{Moggi-Computationallambda}
E.~Moggi.
\newblock Computational lambda-calculus and monads.
\newblock In {\em Proceedings of the Fourth Annual IEEE Symposium on Logic in
  Computer Science (LICS 1989)}, pages 14--23. IEEE Computer Society Press,
  June 1989.

\bibitem{NPP05:CMTT}
A.~Nanevski, F.~Pfenning, and B.~Pientka.
\newblock {Contextual Modal Type Theory}.
\newblock {\em ACM Transactions on Computational Logic}, 9(3), 2008.

\bibitem{pfenning1999system}
F.~Pfenning and C.~Sch{\"u}rmann.
\newblock System description: Twelf -- a meta-logical framework for deductive
  systems.
\newblock In {\em Proc. of CADE}, volume 1632 of {\em Lecture Notes in Computer
  Science}, pages 202--206. Springer-Verlag, 1999.

\bibitem{Pientka08:DependentBeluga}
B.~Pientka and J.~Dunfield.
\newblock Programming with proofs and explicit contexts.
\newblock In {\em PPDP'08}, pages 163--173. ACM, 2008.

\bibitem{belugasys}
B.~Pientka and J.~Dunfield.
\newblock Beluga: A framework for programming and reasoning with deductive
  systems (system description).
\newblock In {\em Automated Reasoning}, volume 6173 of {\em Lecture Notes in
  Computer Science}, pages 15--21. Springer-Verlag, 2010.

\bibitem{prawitz}
D.~Prawitz.
\newblock {\em {Natural Deduction. A Proof Theoretical Study}}.
\newblock {Almqvist~Wiksell, Stockholm}, 1965.

\bibitem{schopenhauer}
A.~Schopenhauer.
\newblock {\em The World as Will and Representation}, volume~2.
\newblock Dover edition, 1966.

\bibitem{schroeder2012honour}
P.~Schroeder-Heister.
\newblock {Paradoxes and Structural Rules}.
\newblock In C.~D. Novaes and O.~T. Hjortland, editors, {\em {Insolubles and
  consequences : essays in honour of Stephen Read}}, pages 203--211. College
  Publications, London, 2012.

\bibitem{schroeder2012proof}
P.~Schroeder-Heister.
\newblock Proof-theoretic semantics, self-contradiction, and the format of
  deductive reasoning.
\newblock {\em Topoi}, 31(1):77--85, 2012.

\bibitem{watkins-02}
K.~Watkins, I.~Cervesato, F.~Pfenning, and D.~Walker.
\newblock {A Concurrent Logical Framework I: Judgments and Properties}.
\newblock Tech. Rep. CMU-CS-02-101, CMU, 2002.

\end{thebibliography}
\end{document}